# Room-temperature colossal magnetoresistance in terraced single-layer graphene

Junxiong Hu, Jian Gou, Ming Yang, Ganesh Ji Omar, Junyou Tan, Shengwei Zeng, Yanpeng Liu, Kun Han, Zhishiuh Lim, Zhen Huang, Andrew Thye Shen Wee, Ariando Ariando*

Dr. J. X. Hu, Dr. G. J. Omar, Dr. S. W. Zeng, Dr. K. Han, Dr. Z. Lim, Dr. Z. Huang, Prof. A. Ariando

NUSNNI, National University of Singapore, Singapore 117411

Email: ariando@nus.edu.sg

Dr. J. X. Hu, Dr. J. Gou, Dr. G. J. Omar, Dr. S. W. Zeng, Dr. K. Han, Dr. Z. Lim, Dr. Z. Huang, Prof. Andrew T. S. Wee, Prof. A. Ariando

Department of Physics, National University of Singapore, Singapore 117542

Dr. J. X. Hu, J. Y. Tan, Prof. Andrew T. S. Wee, Prof. A. Ariando

Centre for Advanced 2D Materials and Graphene Research Centre, National University of Singapore, Singapore 117551

Dr. M. Yang

Institute of Materials Research & Engineering, A*STAR (Agency for Science, Technology and Research), Singapore 138634

Prof. Y. P. Liu

State Key Laboratory of Mechanics and Control of Mechanical Structures, MOE Key Laboratory

for Intelligent Nano Materials and Devices and Institute of Nano Science, Nanjing University of Aeronautics and Astronautics, Nanjing, China 210016

**Keywords**

2D materials, Complex oxides, Interfacial coupling, Disorder, Magnetoresistance

Disorder-induced magnetoresistance (MR) effect is quadratic at low perpendicular magnetic fields and linear at high fields. This effect is technologically appealing, especially in the two-dimensional (2D) materials such as graphene, since it offers potential applications in magnetic sensors with nanoscale spatial resolution. However, it is a great challenge to realize a graphene magnetic sensor based on this effect because of the difficulty in controlling the spatial distribution of disorder and enhancing the MR sensitivity in the single-layer regime. Here, we report a room-temperature colossal MR of up to 5,000% at 9 T in terraced single-layer graphene. By laminating single-layer graphene on a terraced substrate, such as $TiO_2$ terminated $SrTiO_3$, we demonstrate a universal one order of magnitude enhancement in the MR compared to conventional single-layer graphene devices. Strikingly, a colossal MR of >1,000% was also achieved in the terraced graphene even at a high carrier density of ~$10^{12}$ $cm^{-2}$. Systematic studies of the MR of single-layer graphene on various oxide- and non-oxide-based terraced surfaces demonstrate that the terraced structure is the dominant factor driving the MR enhancement. Our results open a new route for tailoring the physical property of 2D materials by engineering the strain through a terraced substrate.

Hybrid two-dimensional (2D) material/complex oxide interface, combining the layered 2D materials and complex oxides, provides various opportunities to study the intriguing physics and develop multifunctional devices, due to the rich properties of both 2D materials and complex oxides, such as 2D electron gas, ferromagnetism, ferroelectricity, and superconductivity.[1,2] The interface coupling between 2D materials and complex oxides will alter the lattice structure and induce the entanglement of the charge, spin and orbital degrees of freedom in 2D materials. This gives rise to a variety of unexpected phenomena, for example, the high dielectric constant $SrTiO_3$(STO) can efficiently screen the long-range Coulomb potential and restore the dominant role of the spin polarization, consequently, resulting in a ferromagnetic phase at the edge states.[3,4] At the graphene-ferroelectric interface, an unusual and robust resistance hysteresis was observed due to the complex surface chemistry of the ferroelectric.[5] When coupled with antiferromagnetic insulators, proximity-induced broken time-reversal symmetry and spin-orbit coupling in graphene make it possible to realize the quantum anomalous Hall effect.[6] However, these emerging phenomena are difficult to realize since the charge transport at the interface is sensitive to the surface environment.[7] In particular, the disorder, like topographic corrugations and charge impurities, will significantly modify the scattering of electrons.[8] Therefore, understanding and controlling the disorder at the interface is crucial to improve the functionality of 2D material-based devices.

Magnetoresistance (MR), the change in the electrical resistance under an external magnetic field, not only provides an efficient tool for probing the fundamental electronic properties in materials, but also results in advanced technological applications.[9] However, intrinsic graphene does not exhibit a MR effect.[10,11] Nevertheless, graphene films can display strong MR by harnessing disorder. According to the classical disorder-induced MR effect, the origin of MR in graphene can be attributed to the fluctuation in carrier density and mobility.[12,13] Various efforts have been devoted to engineering disorder in single-layer graphene, such as by decorating with gold nanoparticles,[14] by fluorination or nitrogen doping,[15,16] or by stacking on a lattice-mismatched substrate.[17] However, the MR of single-layer graphene has yielded limited success thus far at room temperature, typically ranging between 60-775% at 9 T (Supplementary Table 1).[12-23] The multilayer graphene can usually exhibit a large MR due to its interlayer effect. For example, six-layer graphene on hexagonal boron nitride (BN) has been reported to show large room-temperature MR of ~ 2,000% at 9 T.[24] However, the precise control of the layer number in multilayer graphene produced by the mass production techniques, such as roll-to-roll (R2R) or batch-to-batch (B2B) methods, remains challenging.[25] Moreover, as previously reported,[12-24] the MR of graphene will rapidly decay to below 200% at a higher carrier density of $10^{12}$ $cm^{-2}$, which means unintentional doping will lead to the loss of MR sensitivity. The fast batch production of single-layer graphene films, on the other hand, raises the prospect of fabricating a robust graphene magnetic sensor based on a single-layer.[26,27]

Here, we realized a room-temperature colossal MR of up to 5,000% at 9 T, by designing terraced single-layer graphene. We found that the colossal MR of the terraced graphene does not rely on the surface termination of the substrate. Both oxide-based ($TiO_2$, $AlO_2$ or $FeO_2$) and non-oxide-based (AlN) surface terminations produced colossal MR effects consistently above 1,000% even at a high carrier density of ~ $10^{12}$ $cm^{-2}$. Inspired by this finding, we artificially fabricated terraces and steps on a commonly used BN substrate and found that the room-temperature MR of graphene on the terraced BN can also be enhanced to more than 1,000%. Combined scanning tunneling microscopy (STM) and density functional theory (DFT) calculation revealed that the topographic corrugations and inhomogeneous charge puddles both result in disorder in the terraced graphene, which significantly enhances the scattering when applying a magnetic field. Finally, we analyzed and fit our MR and Hall data using the self-consistent effective medium theory.

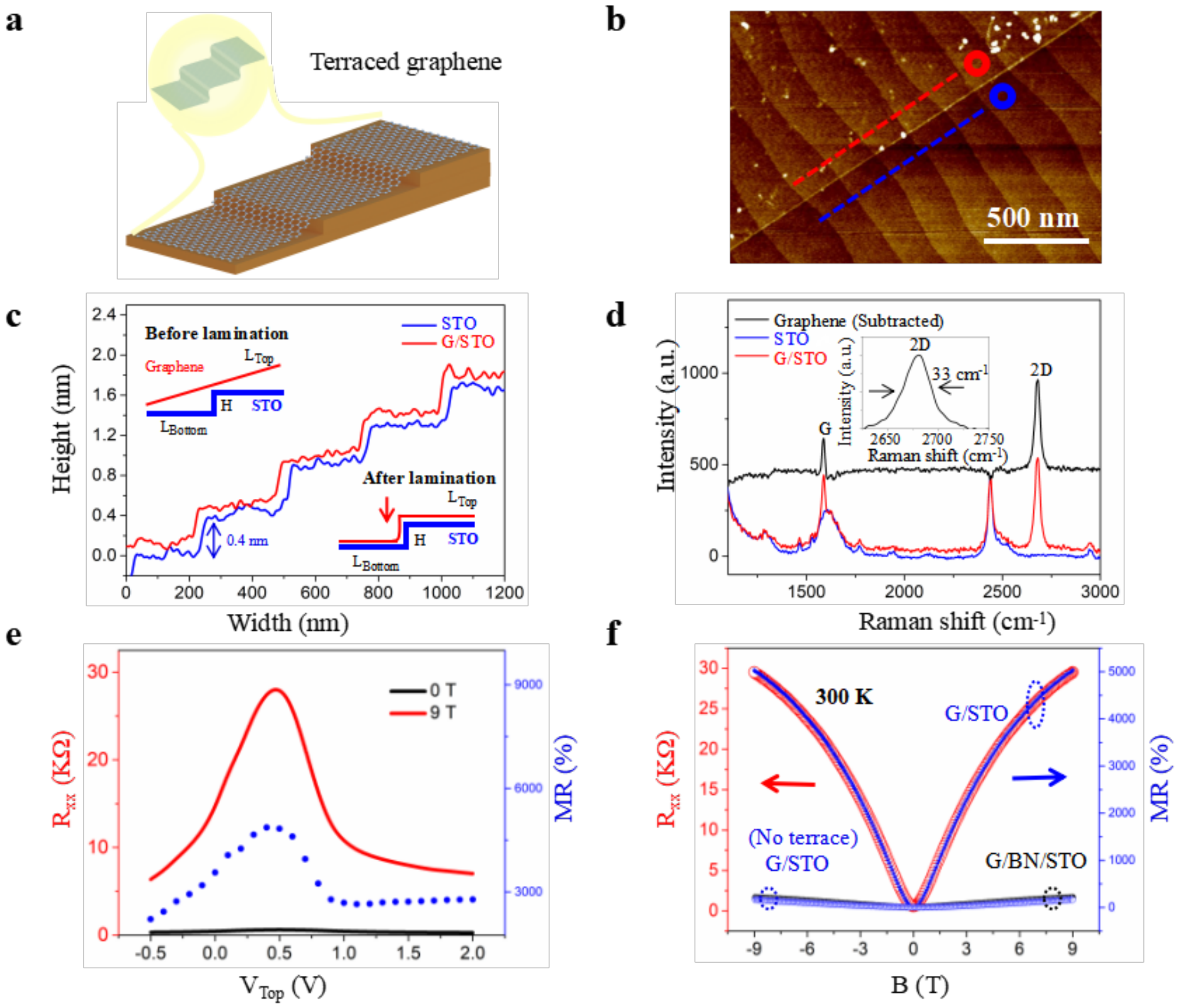

**Figure 1.** Design and characteristics of terraced single-layer graphene. a) A schematic showing the concept of the terraced single-layer graphene formation. Because of the flexibility of graphene, single-layer graphene stacked on a terraced substrate will replicate a similar terrace morphology as the substrate. b) AFM image showing graphene on STO (upper left) and bare STO substrate (bottom right). Both G/STO and STO exhibit atomically flat terraces and steps. c) The surface height profiles taken along the dashed lines indicated in the image in b. The width of the terraces is ~200 nm and the step height is ~0.4 nm. The insets illustrate the lamination process of graphene onto a substrate with an atomic step. d) Raman spectra of G/STO (red line) and STO (blue line) taken at the areas indicated by the circles in b. The inset shows the Raman 2D mode of graphene. e) The longitudinal resistance $R_{xx}$ and as a function of top-gate voltage at 300 K for zero (black line) and 9 T (red line) perpendicular magnetic field. The MR as a function of top-gate voltage at 300 K ( blue solid circles). f) $R_{xx}$ and MR as a function of a magnetic field near the CNP for G/STO with terraces (red and blue lines), G/STO without terrace (blue dashed line), and G/BN/STO (black dashed line).

The terraced single-layer graphene is formed by stacking single-layer graphene on an atomically terraced substrate (Fig. 1a). After the lamination of the graphene on the terraced substrate (Supplementary Fig. S1), the van der Waals (vdW) interaction between the graphene layer and the substrate is expected to be enhanced due to the elastic deformation of graphene, inducing strain both in the terraces and steps.[28] To realize the terraced single-layer graphene, a perovskite-oxide single crystal, $SrTiO_3$ (STO), is utilized as a prototypical substrate for various following reasons. Firstly, atomically flat terraces and steps can be robustly reproduced on the STO surface by selectively etching the SrO layer, leading to a $TiO_2$ terminated surface with a unit-cell (~0.4 nm) step height (see Methods).[29,30] Secondly, STO can induce an inhomogeneous topographical corrugation in graphene due to a large lattice mismatch between graphene and STO, which also contributes to the disorder in graphene.[31] Thirdly, STO has a large dielectric constant at room temperature ($\kappa \approx 200 - 300$) with a wide bandgap of 3.2 eV, providing a strong dielectric screening of charged impurity scattering to ensure the high mobility in graphene is maintained.[4,32] Finally, the possibility of (epitaxially) growing various other materials on STO allows us to examine the influence of surface terminations on the MR.[33]

To fabricate terraced single-layer graphene, graphene flakes were first exfoliated onto a PMGI/PMMA coated silicon substrate. After identification by Raman spectroscopy, single-layer graphene was transferred onto the terraced STO substrate using the dry-transfer technique (see Methods). After the lamination of graphene on the terraced STO, the surface of graphene showed similar terrace and sharp-step features as those of STO (Fig. 1b). To estimate the localized strain induced by the atomic steps, we probed the surface of graphene on STO (G/STO) and the bare STO by line profile scanning using atomic force microscopy (AFM) (Fig. 1c). Similar to the bare STO surface, the graphene surface also showed atomically flat terraces with sharp steps of around 0.4 nm, indicating that the graphene layer has adapted a terraced structure. Because of the elastic deformation, a tensile strain of up to 0.2% can be induced in the steps of the terraced graphene, which can be confirmed by analyzing the Raman 2D mode of single-layer graphene (Supplementary Fig. S2).[34,35] Under elastic deformation, the applied stress $\sigma$ is proportional to the strain $\epsilon$, $\sigma = Y\epsilon$, where $Y = 1 \times 10^9$ Pa denotes Young's modulus of the PMMA film,[34] and thus the PMMA/Terraced graphene is expected to receive applied stress of $\sim 2 \times 10^6$ Pa, enhancing the vdW interaction between the terraced graphene and STO. As shown in the height histogram of G/STO (Supplementary Fig. S3), the average height difference between graphene and STO is only 0.44 nm, indicating single-layer graphene with a narrow vdW gap of $\sim 1$ Å.[36] Subtracting the Raman spectra of G/STO from the bare STO leads to a single

Lorentzian peak at 2,680 cm$^{-1}$ (2D mode), which is 2.2 times of the sharp peak at 1,588 cm$^{-1}$ (G-mode), further confirming the single layer of the terraced graphene (Fig. 1d).[26]

To explore the MR effect in the terraced single-layer graphene, we first investigated the MR of currents traveling either parallel or perpendicular to the steps using standard Hall bar devices (see Methods and Supplementary Fig. S4). Since the MR perpendicular to the steps (1,050%) is slightly larger than the MR parallel to the steps (800%), we focus our study on the case of current perpendicular to the steps. Figure 1e shows the top-gate dependence of the longitudinal resistance ($R_{xx}$) at 300 K. At zero magnetic field, a maximum resistance can be observed at 0.5 V which corresponds to the charge neutrality point (CNP) and indicates p-doping of the terraced graphene.[17,24] Applying a magnetic field of 9 T perpendicular to the sample surface results in a significant increase of $R_{xx}$ for all ranges of the top gate voltage. To quantify the MR effect, we follow the conventional definition that $MR = 100\% \times [R_{xx}(B) - R_{xx}(0)]/R_{xx}(0)$, where $R_{xx}(B)$ and $R_{xx}(0)$ are the resistance at $B$ and zero fields, respectively. The observed MR typically peaks at the CNP and gradually decreases while increasing or decreasing the gate voltage away from the CNP, as shown in the blue curve of Fig. 1e. This gate tunable MR feature is consistent with the previous reports[14,17,24] and can be qualitatively understood by the two-fluid model.[37] Figure 1f shows $R_{xx}$ and MR as a function of a magnetic field ($B$) near the CNP. As the magnetic field is increased, $R_{xx}$ initially shows a quadratic increase in low fields, subsequently turns into a linear increase at moderate fields, and finally saturates at 9 T (Supplementary Fig. S5). The crossover from the quadratic to linear MR behavior is the hallmark of the classical disorder-induced MR.[38] Remarkably, the MR of the terraced single-layer graphene at the CNP is found to be as high as 5,000% at 9 T, which is one order of magnitude higher than that reported on previous single-layer graphene devices at the same conditions (Supplementary Table 1).[12-24]

To confirm that the colossal MR is due to the formation of a terraced structure in graphene induced by the surface topography of the substrate, graphene was laminated on either non-terraced STO surface or BN-buffered terraced STO. First, placing graphene on the as-received STO (without terraces and steps) resulted in a low MR of ~170% at 9 T and 300 K (Supplementary Fig. S6), similar to the MR of graphene on $SiO_2$.[17-19] A clear quantum oscillation was observed at low temperatures with nearly zero MR at low fields, resembling graphene on a flat surface and emphasizing the absence of the substrate corrugation effect. Second, introducing a thick BN layer of 20-30 nm as a buffer between graphene and the terraced STO, thus the steps of the terraced STO can be completely screened by BN, led to a low MR of ~215% (Supplementary Fig. S7), in agreement with the previous report of single-layer graphene on BN.[20] These results demonstrate that without the steps and terraces, the composition (STO, $SiO_2$, and BN) of the surface does not contribute to the enhancement of the MR. Only the surface topography, in particular the atomic steps in STO that enhances the MR in single-layer graphene.[39]

Next, we study how the surface termination influences the colossal MR of the terraced graphene by investigating a range of oxide surface terminations including $TiO_2$-terminated $BaTiO_3$, $AlO_2$-terminated $LaAlO_3$ and $FeO_2$-terminated $LaFeO_3$ (Fig.

2a-c). To verify whether an oxide termination is necessary for obtaining the colossal MR, we also study a non-oxide terminated surface AlN (Fig. 2d). All of the surface terminations were prepared by growing the respective thin film using pulsed laser deposition (PLD) on a terraced STO substrate.[33] Before the thin film growth, the presence of the terraces and steps on the STO substrates was verified by atomic force microscopy (AFM). Six unit cells of the oxide films were then grown on the STO substrate and monitored by reflection high-energy electron diffraction (RHEED) (Fig. 2e). Both the periodic RHEED oscillation and streaky RHEED pattern before and after growth demonstrated the layer-by-layer growth of high-quality thin films. Single-layer

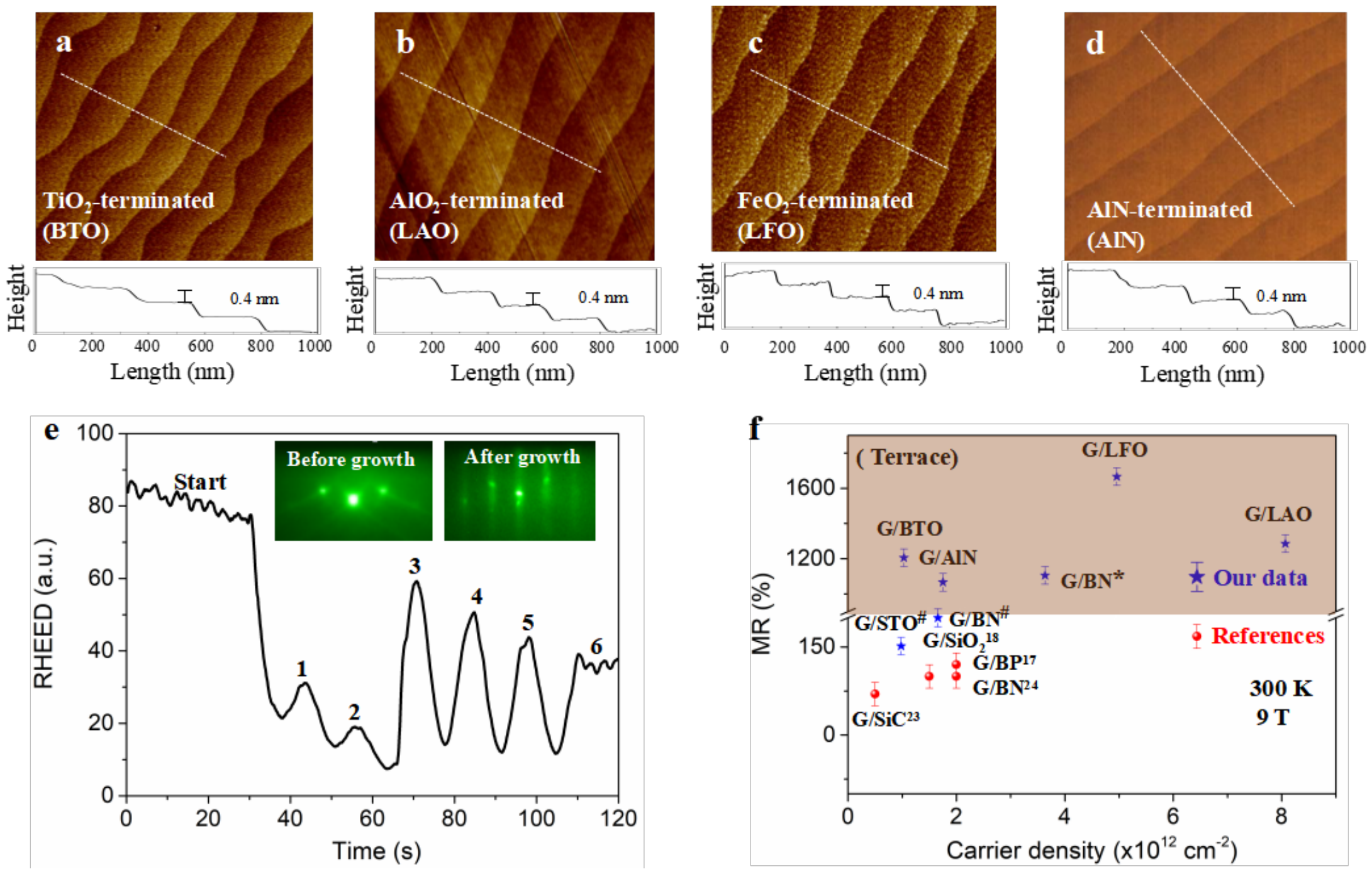

**Figure 2.** MR performance of terraced single-layer graphene on various surface terminations. a-d) AFM images of BTO, LAO, LFO, and AlN grown on a terraced STO(001) substrate. Owing to the layer-by-layer growth mode, the surface is $TiO_2$, $AlO_2$, $FeO_2$, and AlN terminated, respectively. e) Typical RHEED during the growth of BTO, LAO, LFO thin films. The insets show the diffraction patterns before and after growth. All the grown oxide films are six unit cells thick. f) MR of terraced single-layer graphene on various terminations. G/BN* refers to the graphene on the terraced BN. G/STO# and G/BN# refer to the graphene on non-terraced STO and BN-buffered terraced STO, respectively. For comparison, the MR values of conventional single-layer

graphene on various substrates from literature are also included in the figure.

graphene was subsequently transferred onto the freshly prepared surface. Strikingly, all surfaces resulted in a colossal MR >1,200% for oxide-based terminations ($TiO_2$, $AlO_2$, or $FeO_2$) at a relatively high carrier density of $n > 1.2 \times 10^{12}$ $cm^{-2}$ (Supplementary Figs. S8-10) and ~1,060% for non-oxide based termination (AlN) at the carrier density of $1.8 \times 10^{12}$ $cm^{-2}$ (Supplementary Fig. S11). For comparison, previous reports of single-layer graphene prepared on SiC, $SiO_2$, BN, and BP showed a very low MR of around 150% at high-level doping of $10^{12}$ $cm^{-2}$.[14-24] These results suggest that the MR of the terraced single-layer graphene is not dependent on the surface terminations but the surface terraces and steps of the substrate. Inspired by this finding, we artificially fabricated the terraces and steps on a commonly used BN substrate and found that the room-temperature MR of graphene on the terraced BN can also be enhanced to 1,100% at 9 T and 300 K, while it is only 160% for the graphene on flat BN (Supplementary Fig. S12), indicating the crucial role of the terraces and steps of BN in enhancing the magnitude of the MR.

To better understand the MR characteristics of the terraced single-layer graphene, we studied the MR changes with gate voltage, temperature and magnetic field direction in the G/STO system (Fig. 3). Figure 3b shows a gate tunable colossal MR in the terraced single-layer graphene. The MR is maximum close to the CNP and gradually decreases with decreasing gate voltage away from the CNP, indicating the intrinsic character of the terraced graphene is the source of the colossal MR.[14] While the MR shows saturation when lowering the carrier density, there is no sign of saturation with the increase in the carrier density. The observation of nonlinear Hall behavior (Figs. 3c,f) can be associated with the coexistence of two densities of carriers.[24] Figure 3e shows temperature-dependent MR near the CNP. At 2.5 K, apart from the Shubnikov-de Haas oscillations, aperiodic oscillations with a much smaller amplitude of around $0.03\ e^2/h$ can also be seen (Supplementary Fig. S13e), consistent with the criteria that suggest the presence of universal conductance fluctuation (UCF) in our terraced graphene samples.[40,41] With the increase in temperature, the oscillations are rapidly suppressed and completely vanish due to the thermal broadening of Landau levels. Interestingly, above 100 K, the slope of the Hall resistance changes sign from negative to positive (Fig. 3f), indicating a change in the dominant carriers from electrons at low temperature to holes at high temperature (Supplementary Fig. S13), as previously also observed in the graphene/LAO/STO system.[41] The temperature-dependent resistance behavior near the CNP (Fig. S13c) indicates the disorder-dominated transport in terraced graphene.[18] Another evidence that the terraced graphene is consistent with the classical MR picture is the angle dependence of MR (Fig. 3d). $R_{xx}$ reaches a maximum value when the magnetic field is normal to the plane (0° and 180°) and a minimum value when the magnetic field is parallel to the plane (90° and 270°), consistent with the classical Lorentz force causing the deflection of the trajectory of the charged particles.

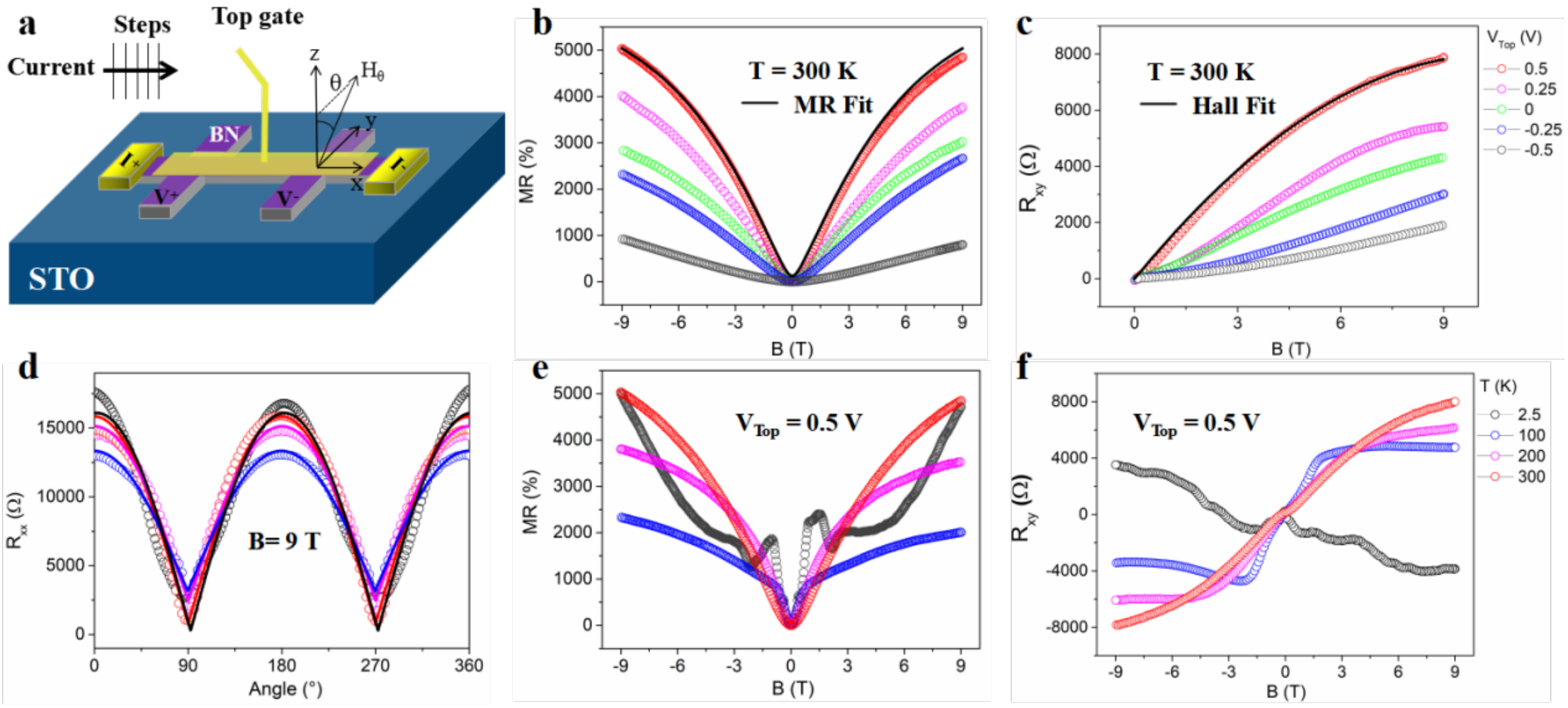

**Figure 3.** Top-gate voltage, temperature and angle-dependent MR in terraced single-layer graphene. a) Schematic illustration of a top-gated terraced single-layer device. The current travels perpendicular to the steps of the terraced graphene. The electrostatic gate voltage is applied to the graphene through a top insulating BN layer, and the external magnetic field is applied perpendicular to the basal plane. b,c) The MR and Hall versus external magnetic field as a function of $V_{Top}$ at 300 K. The MR and Hall near the CNP (close to 0.5 V) are fitted by the effective medium theory. d) The angle dependence of the MR at different temperatures and 9 T, the dots are experiment data and lines are fitting lines, with a relation |cosθ|. e,f) Temperature-dependence of MR and Hall *versus* external magnetic field near the CNP.

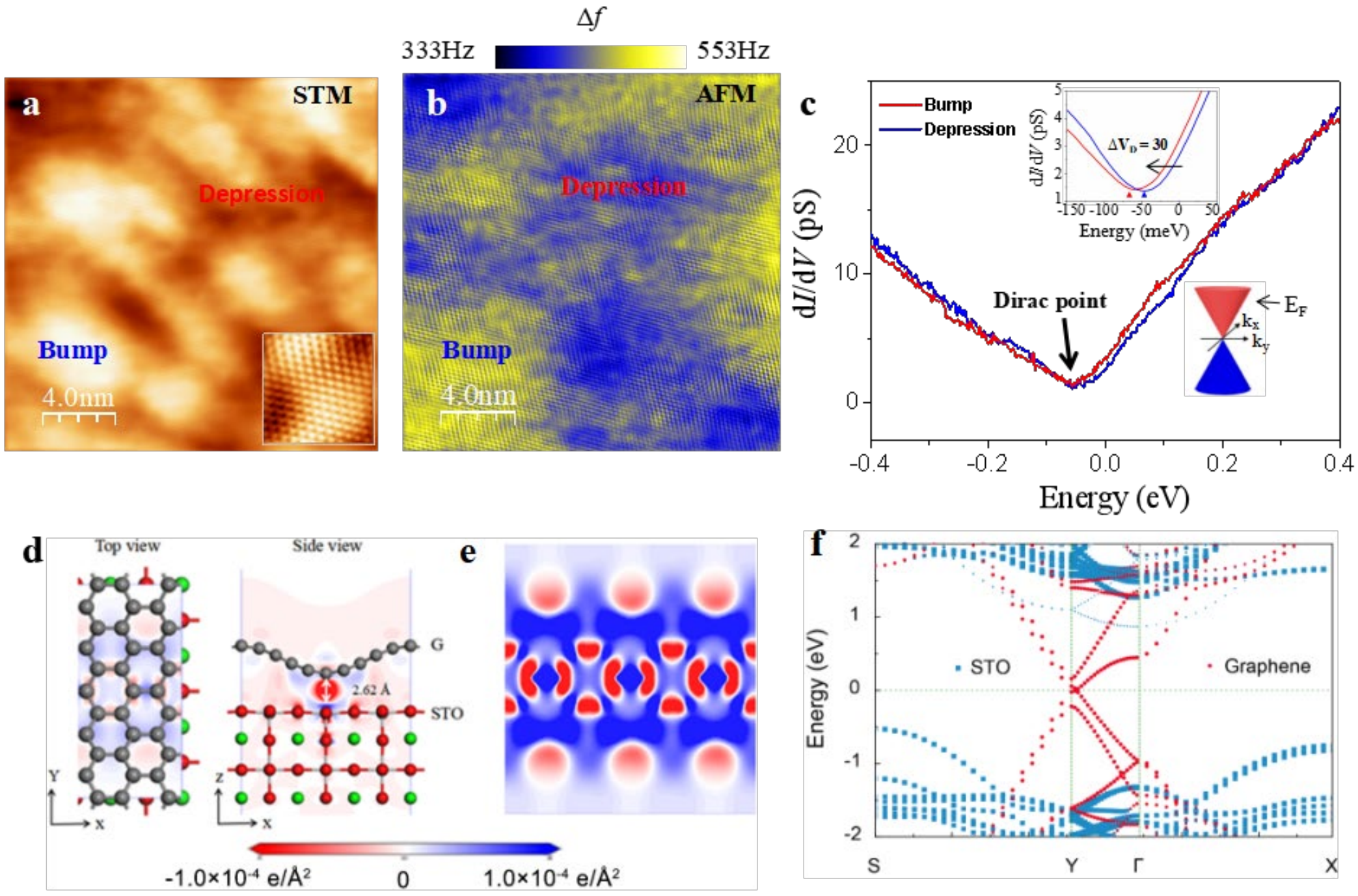

**Figure 4.** Atomic and electronic structure of terraced single-layer graphene. a) STM topography image of graphene on STO (setpoint: $V_s$ = -64 mV, $I_s$ = 200 pA). Inset: STM image with atomic resolution (setpoint: $V_s$ = -300 mV, $I_s$ = 40 pA). b) Nc-AFM image of the same region of graphene as in (a), showing correlated surface corrugations over the whole area. c) Differential conductance (d$I$/d$V$) at both regions of depression (red) and bump (blue) show Dirac-like V-shape spectra. Inset: Zoom-in of the shifted Dirac points. d) Top view (left) and side view (right) of optimized graphene on STO where the superimposed differential charge density was projected along (001) and (010) plane cutting through the lowest and central C atom, respectively. The geometry of graphene becomes warped and the interlayer distance between the bottom of strained graphene and STO is 2.62 Å. e) Top view of differential charge density of graphene/STO projected on strained graphene plane through the lowest C atom which is visualized with a smaller iso-surface value of $1\times10^{-4}$ eÅ$^{-2}$ to highlight the charge density fluctuation. f) Projected band structure of the strained graphene on STO where the red dots and blue square lattices denote the contribution of graphene and STO, respectively. The Dirac cone of graphene still exists despite the large geometry distortion.

Finally, to study the localized disorder of graphene on STO at the nanoscale, we conducted scanning tunneling microscopy (STM) and non-contact atomic force microscopy (nc-AFM) studies of graphene on STO. STM topography imaging shows a random corrugation with bumps and depression regions (Fig. 4a). The atomic-scale image (inset) shows the graphene honeycomb lattice with lattice distortion. The derived surface strain can be as large as 1.7%, as calculated from the measured displacement in the graphene lattices (Supplementary Fig. S14). A comparison between nc-AFM and STM topography over the same area yields apparent correlation (Fig. 4b) between topographic corrugation and electronic fluctuation, indicating that the strain-induced topographic corrugation is the dominant contributing factor for the observed surface fluctuation. Moreover, the shifting of the Dirac points, nearly 30 meV, at bump and depression area indicates that there are inhomogeneous charge puddles on the surface (Fig. 4c), which can be further verified by the d$I$/d$V$ mapping near the Dirac point.[42,43] The d$I$/d$V$ maps at a bias below and above Dirac point showing a nearly inversed spatial density of states (DOS) distribution indicate the different shifting of the Dirac point at different spatial areas with inhomogeneous charge puddles (Supplementary Fig. S15). As a result, graphene on STO contains both disorders of topographic corrugations and inhomogeneous charge puddles at the terrace, which significantly enhance the electron scattering in a magnetic field.

To understand the origin of the localized strain in the terraced graphene, we performed DFT calculations to gain insights into the interfacial interaction between graphene and STO (see Methods). Due to a large mismatch of the lattice constants between graphene and STO, the graphene may form incommensurate superlattice structures on STO, in which the different regions of graphene experience variable strains from the STO substrate. Therefore, in the simulation, we considered two different manually strained graphene supercells on the STO substrate (see the details in Methods). As Fig. 4d and e show, the optimized structure shows that graphene on STO with the larger compressive strain (~6.9%) becomes warped. The highly strained graphene results in a pronounced charge inhomogeneity in the graphene (Fig. 4e). However, the electronic structure of graphene does not change significantly (Fig. 4f) since the Dirac cone of graphene can still be seen in the band structure,[44] consistent with the $dI/dV$ curve measured by STM. For graphene on the BN substrate, the optimized structure of graphene is flat and the charge density redistribution is uniform (Supplementary Fig. S16) due to the matched lattice constants between graphene and BN. Moreover, the interfacial interaction of G/STO is stronger than that of G/BN and G/BP, as evidenced by the larger adsorption energy, a shorter interfacial spacing, and more pronounced charge redistribution (Supplementary Fig. S17). The stronger interaction and shorter spacing between graphene and STO facilitate the effective coupling between graphene and STO, leading to stronger scattering due to charge puddles in graphene and phonon scattering from the substrate. On the other hand, the preserved Dirac cone of graphene on the STO substrate indicates that the carrier mobility of strained graphene does not degrade significantly. All these may contribute to the observed colossal MR in the terraced graphene.

Various theoretical models have been proposed to understand the disorder-induced MR effect. The two-fluid model assumes the coexistence of electrons and holes at the CNP that can explain the sharp MR peak at the CNP but with a saturating MR feature in high fields.[37] The model also predicts that the MR approaches zero far away from the CNP, which is not consistent with our observation. The other two models that have successfully described the disorder-induced MR are the random resistor network model and the self-consistent effective medium theory,[45,46] which have been demonstrated to belong to the same universality class.[38] The effective medium theory considers the presence of electron-hole puddles and the strong distortion of current lines at the boundaries between n and p-type puddles under a magnetic field. Based on the effective medium theory, we successfully fit the MR data near the CNP (Fig. 3b), by ensuring that the fraction of electron puddles is 0.48 and n = $1.6\times10^{15}$ $m^{-2}$; p = $1.12\times10^{15}$ $m^{-2}$; $\mu_n$ = 2.15 $m^2$/V; $\mu_p$ = 2.71 $m^2$/Vs and fit the Hall data (Fig. 3c) by ensuring that $n$ = $2.3\times10^{15}$ $m^{-2}$; $p$ = $4.5\times10^{15}$ $m^{-2}$; $\mu_n$ = 3.68 $m^2$/Vs; $\mu_p$ = 0.8 $m^2$/Vs. Here different parameters are used to fit the MR and Hall data, and this discrepancy may be attributed to the assumptions made in this model (Supplementary Fig. S18).

The band structures of 2D semiconductors like transition metal dichalcogenides are usually sensitive to the strain. For example, the bandgap of $MoS_2$ can be modulated by the tensile strain, leading to excellent electrical and optical performance.[47] Our STM and DFT calculations (Figs. 4c,f) on the contrary show that the electronic structure of terraced graphene does not change significantly with the application of strain, resulting in a robust carrier mobility. In a disordered system, this high mobility is expected to contribute to the large MR.[46] Note that our MR in terraced graphene based on standard Hall bar structure is the physical MR, not from the geometrical effect, since either the physical or geometrical contribution may dominate the observed MR. The MR can be further enhanced by the geometrical effect, like extraordinary magnetoresistance (EMR).[48] Moreover, not only can the strain enhance the MR effect in terraced graphene, but it may also lead to the pseudo-magnetic field (PMF). This is because a non-uniform strain can deform the corners of the graphene Brillouin zone, shifting the Dirac cones at K and K' points in the opposite directions, creating PMF.[49] The strain-induced large MR and PMF effects have been previously demonstrated in G/BP system.[17,31] We believe our terraced graphene systems, apart from the MR effect, will also exhibit the PMF effect which would be an interesting future study. Recent electrical transport and STM studies of PMF in terraced graphene indicate that our proposal is plausible.[50,51]

In conclusion, we designed and demonstrated a viable way to artificially engineer disorder in single-layer graphene by laminating it on a terraced substrate. This results in a strained and terrace-like single graphene layer that exhibits a colossal MR of one order of magnitude larger than that of conventional flat graphene at room temperature. The MR of the terraced graphene is gate tunable, nearly temperature independent, and magnetic field angle-dependent following the classic disorder-induced MR effect. Unlike the enhanced MR effect previously obtained by chemical doping or surface decoration of the graphene layer which modifies the electronic band structure of graphene, the disorder-induced MR by substrate terraces is highly reliable, reproducible, and non-destructive. The intrinsic property of graphene is well maintained. Most importantly, the MR sensitivity remains robust even at a high-doping level of $10^{12}$ $cm^{-1}$. Our methods open a new route for tailoring the physical property of 2D materials, especially for those strain-sensitive 2D semiconductors such as $MoS_2$,[47] $MoTe_2$ or 2D magnet, $Cr_2Ge_2Se_6$.[52,53] Many novel phenomena are waiting to be revealed in such kind of terraced 2D materials.

**Experimental Section**

*HF treatment of STO(001):* To obtain the atomically flat terraces and steps in STO, the as-received (001)-oriented STO single crystalline substrate (CrysTec GmbH) was first ultrasonically soaked in analytical grade and demineralized water for 10 mins to form Sr-hydroxide complex, followed by a 30 s dipping in standard and commercially available buffered HF solution. The chemical reactions are: $SrTiO_3 + H_2O = Sr(OH)_2 + TiO_2$, $Sr(OH)_2 + 2HF = SrF_2 + 2H_2O$. During the process, the SrO layer will be chemically selectively dissolved by HF, ensuring that the surface is purely $TiO_2$-terminated with atomically-flat terraces and steps. To remove the remnants of the previous treatments and facilitate recrystallization, a final annealing step was performed at 950 °C for 2 hours. The surface topography of the terraced STO was characterized by atomic force microscopy (AFM) (Bruker Icon3, Bruker Corporation, USA) with Tespa-V2 AFM cantilevers in tapping mode.

*Epitaxial growth of perovskite oxides:* Epitaxial BTO, LAO, and LFO thin films were grown on $TiO_2$-terminated (100)-oriented STO substrates using pulsed laser deposition (PLD) system at 750 °C. An excimer (KrF) laser with a wavelength of 248 nm, an energy fluence of 0.8 J $cm^{-2}$, and a repetition rate of 2 Hz was used. BTO, LAO, and LFO thin films were deposited at an oxygen partial pressure of $10^{-3}$ Torr. To minimize the oxygen vacancy formation, all samples were subsequently post-annealed at 600 °C in a high oxygen environment of 200 Torr for 1 h and then cooled down to room temperature. During the deposition, an in situ reflection high energy electron diffraction (RHEED) was used to monitor the thickness of BTO, LAO, LFO thin films. The amorphous AlN film (5 nm) was grown under $4\times10^{-5}$ Torr background pressure at room temperature using 750 laser pulses with an energy fluence of 0.8 J $cm^{-2}$ and a repetition rate of 10 Hz and subsequently annealed at 550 °C for 2 hours to remove oxygen vacancy in STO.

*Device fabrication and characterization:* To fabricate a terraced single-layer graphene device, natural graphite (NGS, Germany) were first mechanically exfoliated on a PMGI/PMMA-coated silicon substrate. After identifying the single-layer graphene by Raman spectroscopy (Renishaw), the PMGI layer was slowly dissolved by the MF319 solution, and subsequently, the Graphene/PMMA bilayer was cleaned thoroughly in DI water. The Graphene/PMMA bilayer was then transferred onto the terraced substrate using the dry-transfer technique. Finally, the PMMA layer was removed by acetone and terraced single-layer graphene was finally obtained. To fabricate a Hall bar channels either parallel or perpendicular to the steps of the substrate, a Au marker (10 nm thick) was deposited around the graphene flake and then annealed at 350 °C for 2 hours under $H_2$/Ar gas to remove the PMMA residue. The angle between the steps and horizontal lines of the Au marker can be clearly identified by AFM. Based on the measured angle, the Cr/Au (2/60 nm) electrodes can be exactly designed and deposited both parallel and perpendicular to the steps using Lesker Thermal Evaporator. Finally, $O_2$ plasma was used to obtain 8μm by 2μm graphene stripes.

*Electrical transport measurement:* The electrical transport measurements were carried out in a physical property measurement system (PPMS) interfaced with a source meter (Model 2400, Keithley Inc.) and a multimeter (Model 2002, Keithley Inc.). A direct current of 1 μA was applied through the sample to measure the longitudinal resistance in a magnetic field (perpendicular to the sample plane) with the magnitude varying from -9 to 9 T. Before the measurement, the samples were annealed for 2 h at 350 K in $H_2$/Ar atmosphere to remove the PMMA residual and any adsorbed water vapor. To apply top-gate bias, the source terminal was connected to the top gate electrode and the leakage current through the BN layer was monitored. At low magnetic fields between 0-1 T, the linear slope $s = dR_{xy}/dB$ determines the carrier density $n = 1/se$. The Hall mobility is calculated through the expression $\mu = 1/neR_{xx}$.

*STM/ nc-AFM imaging:* STM/nc-AFM measurements were carried out in a LT-Omicron STM system (base pressure $1\times10^{-10}$ mbar). G/STO device was degassed to 600 K to get rid of surface contamination and then transferred into the STM head to cool down to 77 K for STM/nc-AFM scanning. qPlus sensor with electrochemically etched tungsten tips was used. Bias voltages were applied to the sample for STM and STS measurements. STS was obtained by the built-in lock-in amplifier in the Nanonis controller, with which a 20 mV (r.m.s), 963Hz sinusoidal modulation was superimposed on the sample bias. For nc-AFM imaging, the constant height mode with an oscillation amplitude of about 2 Å was used to record the frequency shift ($\Delta f$). All the STM/nc-AFM images were processed using WSxM software.

*Theoretical calculations:* All calculations were carried out using density-functional theory (DFT) based on Vienna ab initio simulation package (VASP5.4.18) with the Perdew–Burke–Ernzerhof (PBE) approximation for the exchange-correlation functional and projector-augmented wave (PAW) potential for the electron-ion interaction. The cut-off energy for the plane-wave expansion was set to 500 eV, and the vacuum layer with a thickness larger than 15 Å was applied in all the calculations. An effective Hubbard $U$ ($U = 4.0$) was applied to $d$ orbital electrons of Ti. For all the hybrid structures, van der Waal interaction was included using the DFT-D3 method. The electronic convergence was set to $1.0\times10^{-5}$ eV, and the force on each atom is optimized smaller than 0.01 eV/Å.

For the interface models of graphene on the STO substrate, we have considered two different graphene supercells on the STO substrates. One is to place $(\sqrt{3} \times 5)$ graphene supercell on the $(1 \times 3)$ STO substrate, in which about 6.9% and 3.2% compressive strain were applied along the armchair and zigzag edge of graphene, respectively (large strain, L-G/STO). Another is based on $(4\sqrt{3} \times 7)$ graphene supercell on the $(3\sqrt{2} \times 3\sqrt{2})$ STO, where the armchair and zigzag edge of graphene were compressed by about 1.3% and 2.3%, respectively (small strain, S-G/STO). For the interface model of graphene on a h-BN substrate, graphene was stretched by about 1.3%. Besides, the interface structure of graphene on the black phosphorus (BP) substrate (G/BP) was constructed by placing strained $(4 \times \sqrt{3})$ graphene supercell on the bi-layer $(3 \times 1)$ BP supercell, with a tensile strain of about 1.1% and 1.6% applied on the zigzag and armchair edge of graphene, respectively. Γ-centered $12 \times 12 \times 1$, $10 \times 10 \times 10$, $8 \times 4 \times 1$, $2 \times 2 \times 1$, and $4 \times 12 \times 1$ $k$-point meshes were used for sampling the Brillouin zones of graphene (graphene/h-BN), STO bulk, L-G/STO, S-G/STO, and G/BP, respectively.

The adsorption energy for graphene on STO, h-BN, and BP was estimated by $\Delta E_{ad} = E_G + E_X - E_{G+X}$, where $E_{G+X}$ is the total energy of the hybrid structures for graphene on STO, BN, or BP, $E_G$ is the total energy of isolated graphene, and $E_X$ is the total energy of isolated STO, BP or BP layers, respectively. The larger $\Delta E_{ad}$ indicates stronger interfacial interaction between graphene and the substrate. The charge density redistribution for graphene on the substrates can be seen from the differential charge density difference as defined by $\Delta\rho = \rho_{G+X} - (\rho_G + \rho_X)$, in which $\rho_{G+X}$ is the charge density of the hybrid structures, and $\rho_G$ and $\rho_X$ is the charge density of isolated graphene and the substrate (STO, h-BN or BP), respectively.

**Acknowledgments**

J. X. H, J. G, M.Y contributed equally to this work. This work is supported the Singapore National Research Foundation (NRF) under the Competitive Research Programs (CRP Grant No. NRF-CRP15-2015-01) and the Agency for Science, Technology, and Research (A*STAR) under its Advanced Manufacturing and Engineering (AME) Individual Research Grant (IRG) (A1983c0034). M.Y. would like to thank the funding support from the Agency for Science, Technology, and Research (A*STAR) 2D PHAROS project (project No: SERC 1527000012). A.A. and J.X. would like to acknowledge the technical support from the Center of Advanced 2D Materials for the device fabrication.

**Competing interests**

The authors declare no competing interests.

**Additional information**

Supplementary information is available for this paper at.

Reprints and permissions information is available at.

Correspondence should be addressed to A. Ariando.

Publisher's note: Springer Nature remains neutral with regard to jurisdictional claims in published maps and institutional affiliations.

**Data availability.** The data that support the plots within this paper and other findings of this study are available from the corresponding authors upon reasonable request.

SUPPLEMENTARY INFORMATION

# Room-temperature colossal magnetoresistance in terraced single-layer graphene

Junxiong Hu, Jian Gou, Ming Yang, Ganesh Ji Omar, Junyou Tan, Shengwei Zeng, Yanpeng Liu, Kun Han, Zhishiuh Lim, Zhen Huang, Andrew Thye Shen Wee, Ariando Ariando*

**Table of Contents**

## S1. Lamination process of single-layer graphene on terraced STO

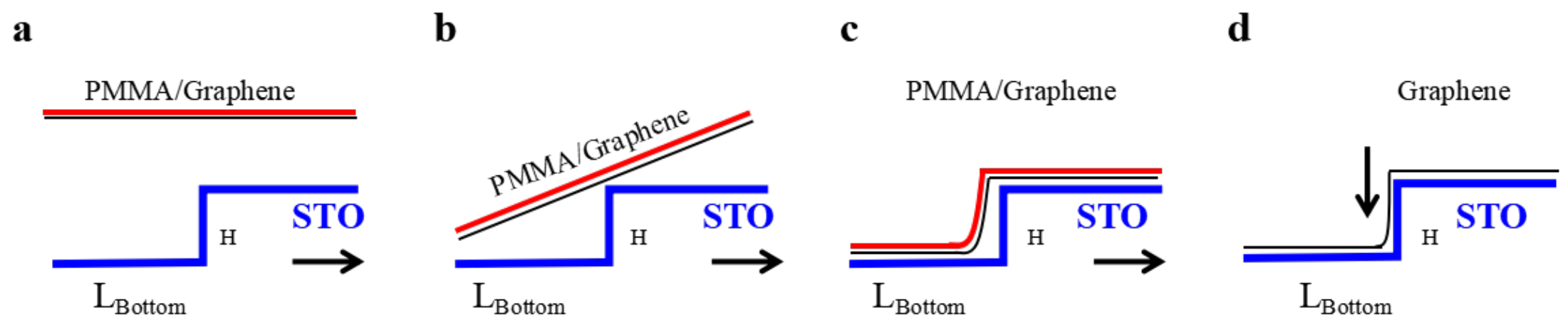

**Supplementary Figure S1 | Lamination process of single-layer graphene on terraced STO.** The lamination process can be summarized into four steps: **a,** Preparation of a free-standing PMMA/graphene film and a terraced substrate. **b,** Transferring the PMMA/graphene film on the terraced substrate. **c,** After the transfer process, the substrate heated to above the PMMA transition temperature (~100 ℃) that causes the graphene layer to change from a hard, glassy state into a rubbery state. This results in a good and uniform contact between graphene and the terraced substrate. The entire film forms a stable vdW interaction with the terraced substrate separated by a vdW gap. **d,** Cooling down to room temperature and subsequently dissolving the PMMA by acetone.

## S2. Strain estimation of terraced single-layer graphene on STO

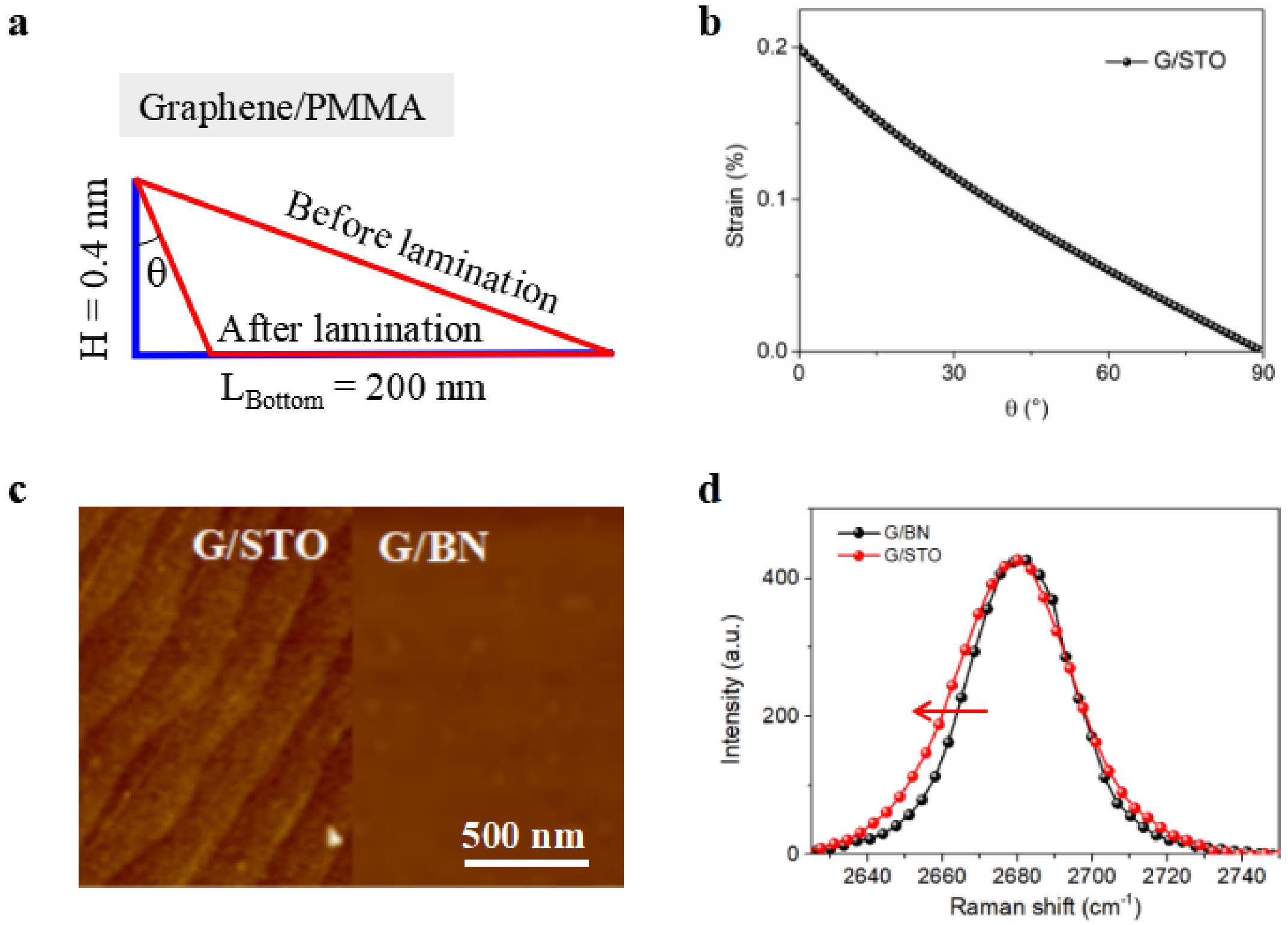

**Supplementary Figure S2 | Strain estimation of terraced single-layer graphene on STO. a,** Model of the lamination process of graphene on a single-step of STO. **b,** Calculated strain dependent with the lamination angle θ. The maximum strain is 0.2% when θ = 0**. c,** AFM images of graphene on STO and BN, respectively. **d,** Raman 2D peak for graphene on STO and BN. The data are normalized by the amplitude of the 2D peak. Symbols are the experimental data, with lines of the corresponding Lorentzian fit in the respective colour. Black spheres and line: graphene on flat BN. Red spheres and line: graphene on terraced STO. The full width at half-maximum of 2D mode becomes widened and the position of the 2D mode shows a red-shift due to the induced strain in G/STO.

To theoretically estimate the tensile strain in terraced single-layer graphene, we establish a lamination model of a PMMA/graphene bilayer onto STO with an atomic step. In this model, three assumptions are made: First, two anchor points are located at the end of every step. Second, the adsorption energy between graphene and STO is strong enough, so the lamination point will attach to either side of the step. Third, L » H, so the lamination point will locate at the length side. Based on these three assumptions, we can get the total length between two anchor points before lamination:

$$L_{Total}^{Before} = \sqrt{L_{Bottom}^2 + H^2} \qquad (S1)$$

After the lamination, the entire film forms a stable vdW interaction with the substrate separated by a vdW gap. The film is strained, and in this scenario, the total length is given by

$$L_{Total}^{\text{After}} \approx \frac{h}{\cos\theta} + l - h\tan\theta \qquad (S2)$$

Then we can get the tensile strain:

$$\varepsilon = \frac{L_{Total}^{\text{After}}\text{-}L_{Total}^{\text{After}}}{L_{Total}^{\text{After}}} \times 100\% = \frac{\frac{h}{\cos\theta}+l-h\tan\theta-\sqrt{l^2+h^2}}{\sqrt{l^2+h^2}} \times 100\% \qquad (S3)$$

where $l = 200$ nm and $h = 0.4$ nm measured by AFM. Then we can get the angle-dependent strain distribution in the terraced graphene, as shown in Fig. S2b. When the angle is zero, the strain reaches its maximum value, 0.2%, and decreases with the increasing angle θ. The strain can be experimentally studied by the Raman 2D mode of graphene. We study the Raman spectrum of graphene on flat BN and terraced STO (Fig. S2c). Position of 2D peak shift from 2,682 $cm^{-1}$ to 2,680 $cm^{-1}$ and the full width of half-maximum also increases from 30 $cm^{-1}$ to 34 $cm^{-1}$, which would correspond to a tensile strain of up to 0.2%, in agreement with our calculation. Moreover, under elastic deformation, the applied stress, σ, is proportional to the strain $\epsilon$ according to

$$\sigma = Y\epsilon \qquad (S4)$$

where Y denotes Young's modulus of the PMMA film, which is $1\times10^9$ Pa. Therefore, the PMMA/terraced graphene bilayer is expected to receive an applied stress of ~$2\times10^6$ Pa. Such a large pressure applied to the film is expected to decrease the vdW gap between the terraced graphene film and STO, as shown in Fig. S3.

## S3. Van der Waals gap between terraced single-layer graphene and STO

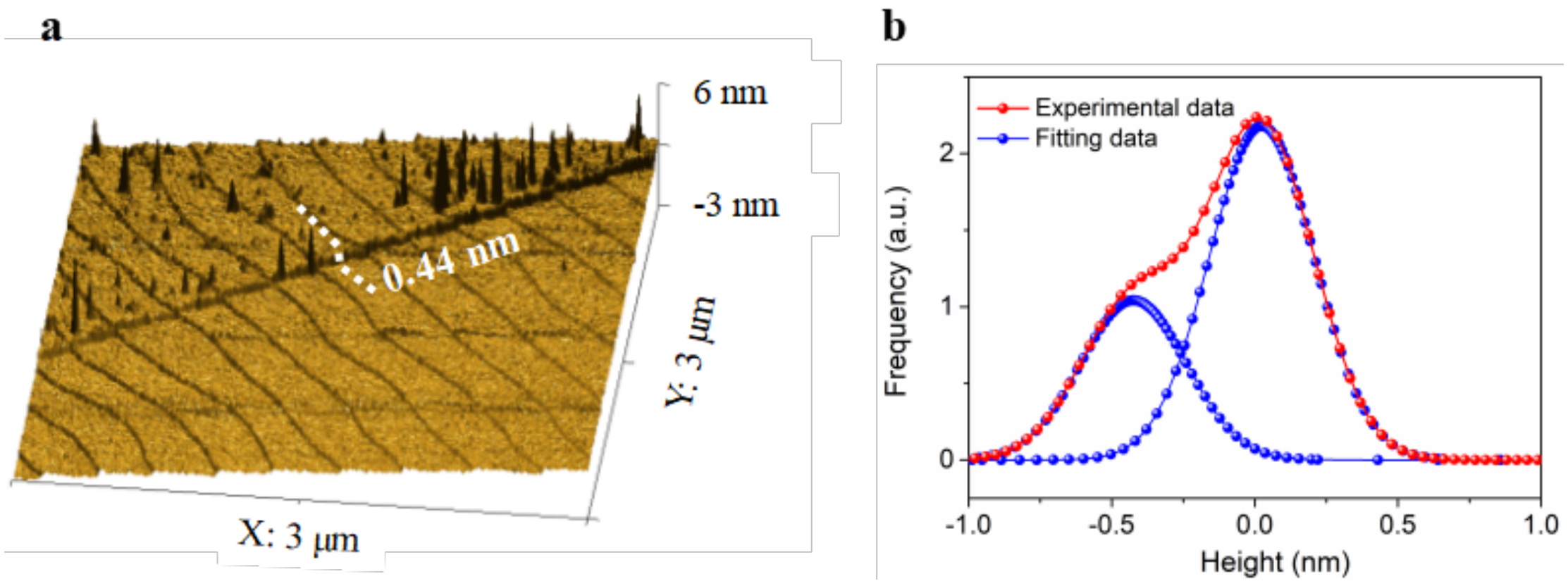

**Supplementary Figure S3 | Histogram of the height distribution of terraced single-layer graphene on STO. a,** 3D AFM image for graphene on STO. The top left side is graphene and the bottom right side is the bare STO. **b,** Histogram of the height distribution of terraced graphene on STO. The red spheres and line are experimental data and the blue spheres and lines are Lorentzian fit.

The 3D AFM image shows the terraced structures of graphene and STO. After the lamination of graphene on the terraced STO, graphene also adapts a terraced structure and the other protrusion is the PMMA residue. It is clear that the height of graphene is higher than the bare STO. The histogram of graphene on STO shows one broad peak with a shoulder, which can be Lorentzian fitted into two single peaks and the distance between the two peaks is 0.44 nm. Considering the thickness of single-layer graphene is 0.34 nm, thus we can induce that the vdW gap between graphene and STO is around 0.1 nm.

## S4. MR of terraced single-layer graphene parallel and perpendicular to the steps

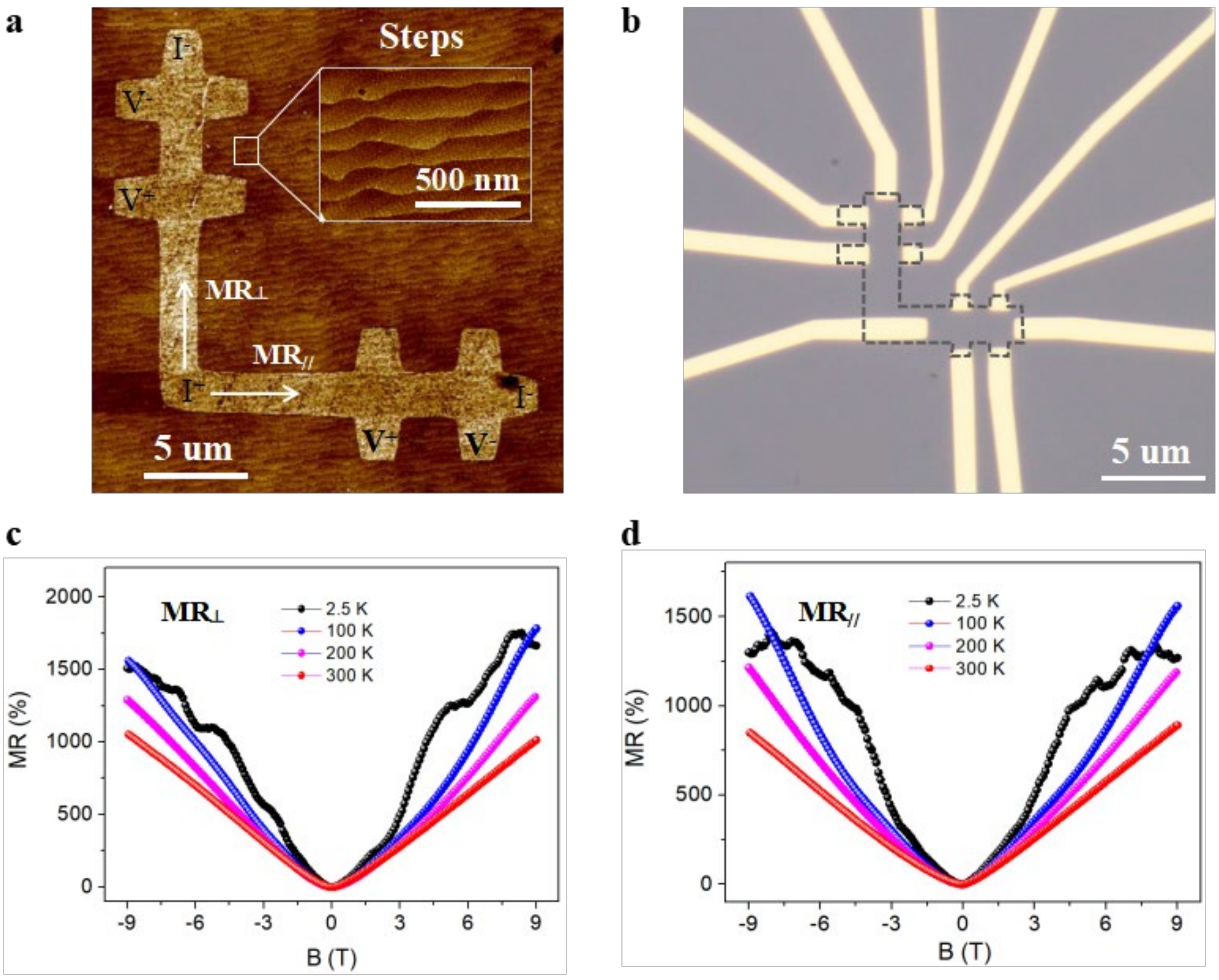

**Supplementary Figure S4 | MR of terraced single-layer graphene parallel and perpendicular to the steps. a,** AFM image of two orthogonal Hall bar structure parallel and perpendicular to the steps. The inset shows the magnified steps of terraced STO. **b,** Optical image of the fabricated terraced single-layer graphene device. The location of graphene is indicated by the dark dashed line. **c, d,** Temperature-dependent MR parallel and perpendicular to the steps.

Two orthogonal Hall-bars are designed to study the MR in which the current is either perpendicular or parallel to the steps (see Methods and Fig. S4a). In the structure depicted in Fig. S4a, the current in horizontal Hall bar is parallel to the steps while the current in upright Hall bar is perpendicular to the steps. Since the two orthogonal Hall-bars are structured in the same graphene layer, thus the carrier density would be similar and their MR effects are comparable. The optical micrograph of the fabricated Hall-bar device is shown in Fig. S4b. The temperature-dependent shows that both the MR are nearly temperature independent at a low magnetic field and show a discrepancy with the increase of the applied magnetic field. At low temperature, the MR shows SdH oscillations and with the increase of temperature, the SdH oscillations disappear because of a thermal broadening of the Landau levels. At 300 K, the MR linearly increases with the magnetic field. At 9 T, the MR of current perpendicular to the steps is 1,050%, slightly larger than the perpendicular to the steps, which is 800%.

**S5. Disorder-induced MR behavior in terraced single-layer graphene**

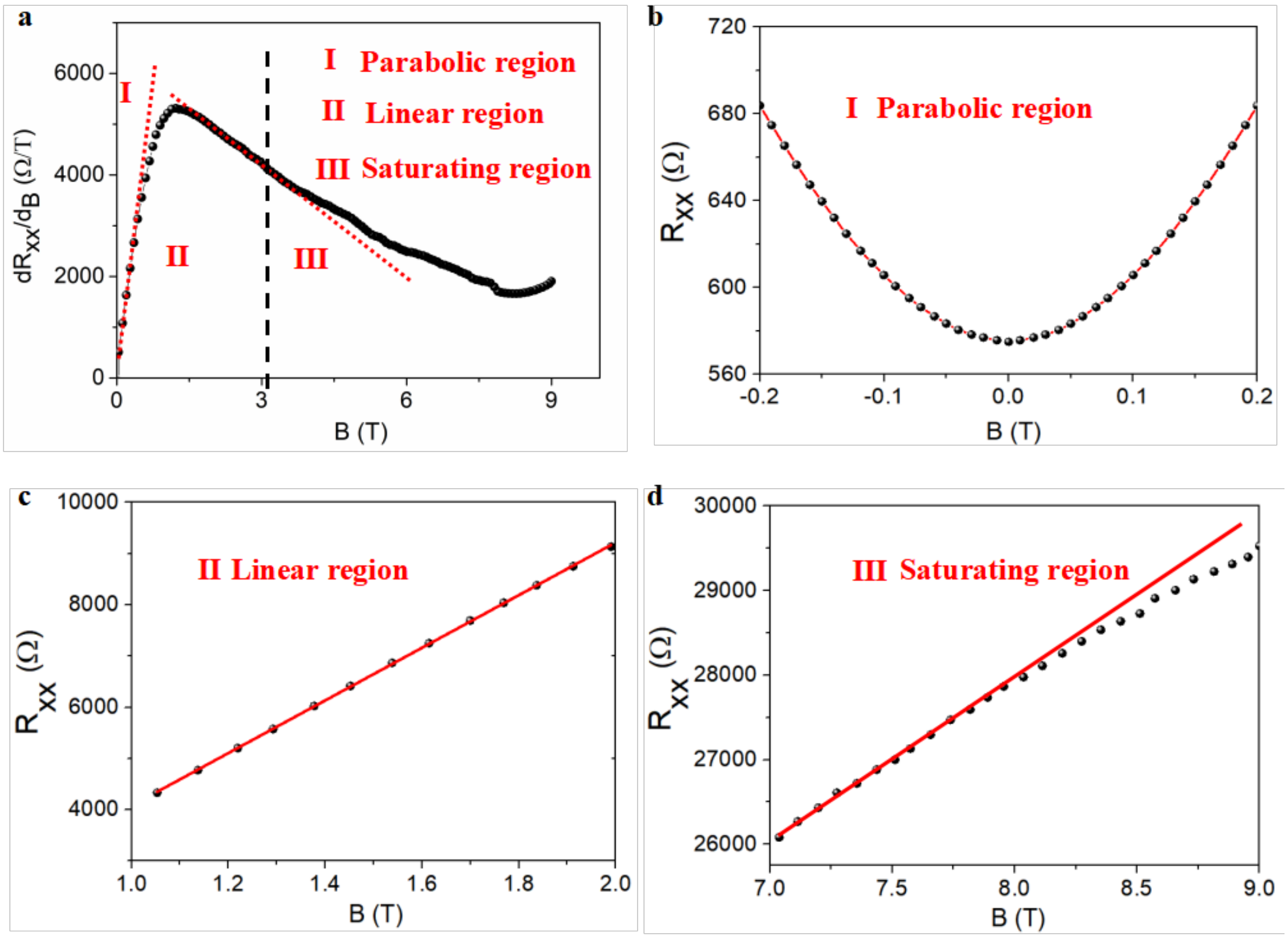

**Supplementary Figure S5 | Three different regions of MR in terraced single-layer graphene. a,** $R_{xx}$ numerical derivative $dR_{xx}/dB$ as a function of B. $dR_{xx}/dB$ shows a maximum at 1.2 T. According to the slope, the figure can be divided into three regions: **b,** At low magnetic fields, MR shows a quadratic B dependence. **c,** In the intermediate B region, a linear MR is observed. **d,** In the high B region, signs of MR saturation are seen. Black spheres are experimental data and red lines are the corresponding fit. The MR value in terraced graphene shows a quadratic dependence in the low field region, linear dependence in the intermediate field region, and saturated behaviour in the high field region. These characteristics are in agreement with the classical disorder-induced MR effect, indicating that the MR of terraced graphene belongs to the classical MR category, which can be understood by the self-consistent effective medium theory, as discussed in Figure S18.

## S6. MR of single-layer graphene on as-received STO

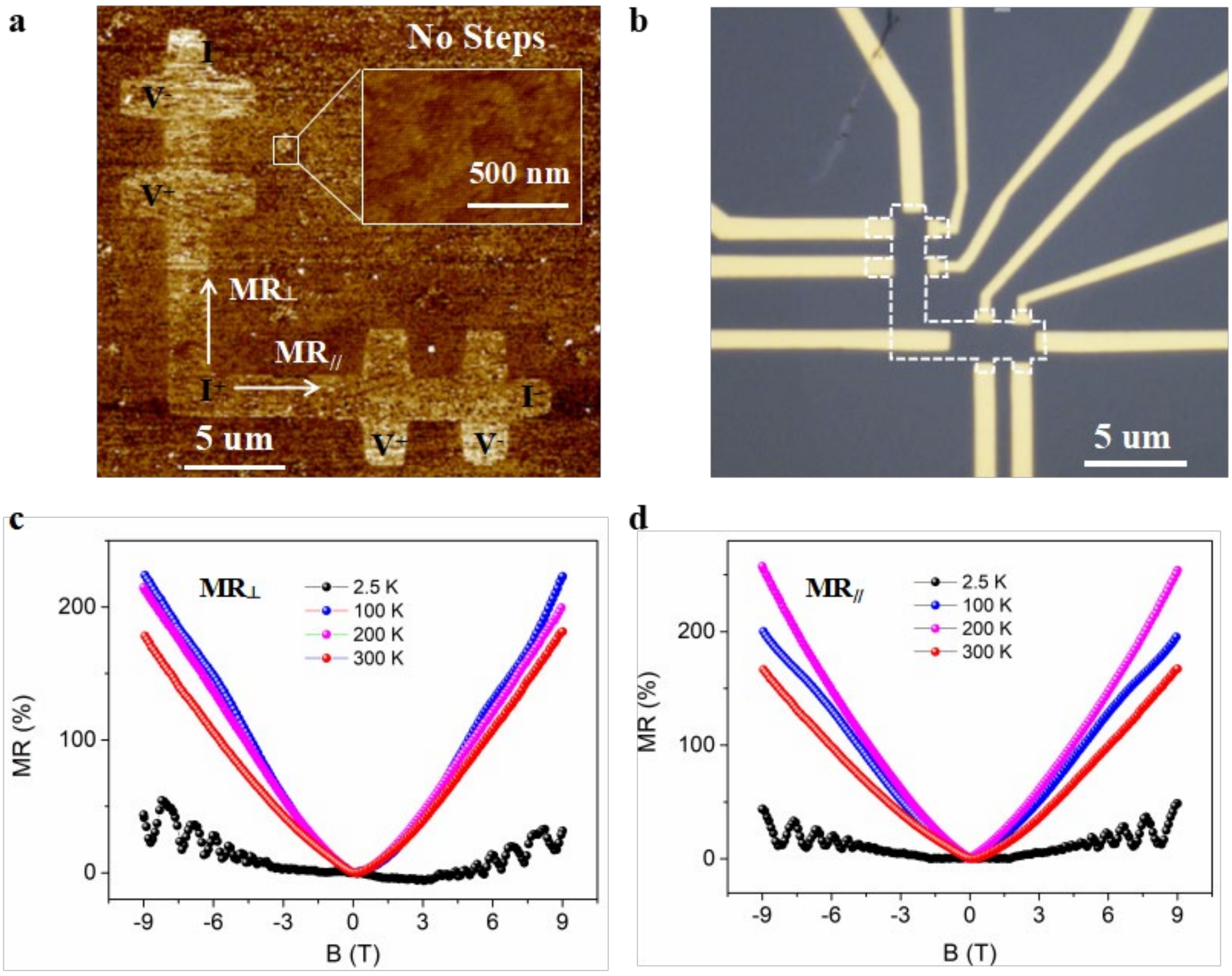

**Supplementary Figure S6 | MR of single-layer graphene on STO without terraces and steps. a,** AFM image of two orthogonal Hall-bar structures. The inset shows the magnified area of as-received STO. Since there is no terrace and steps in the as-received STO, the two orthogonal Hall-bars are designed in a random direction. **b,** Optical image of the fabricated single-layer graphene device. The location of graphene is indicated by the white dashed line. **c,d,** Temperature-dependent parallel and perpendicular MR. Temperature-dependent MR both show the quantum oscillation at low temperature, similar to previously reported graphene on $SiO_2$, indicating the high quality of graphene on as-received STO. At the low magnetic field, the MR is nearly zero, showing the intrinsic property of graphene. With the increase of the magnetic field, the quantum oscillation becomes significant. At 300 K, the quantum oscillation disappear and a positive MR value of 220% both observed in parallel and perpendicular MR. Compared with the large MR of 1050% for graphene on terraced STO (Fig. S4), the MR of graphene on as-received STO is rather small, indicating the crucial role of terraces and steps of terraced STO in enhancing MR.

**S7. Comparison of MR of single-layer graphene on flat BN and terraced STO**

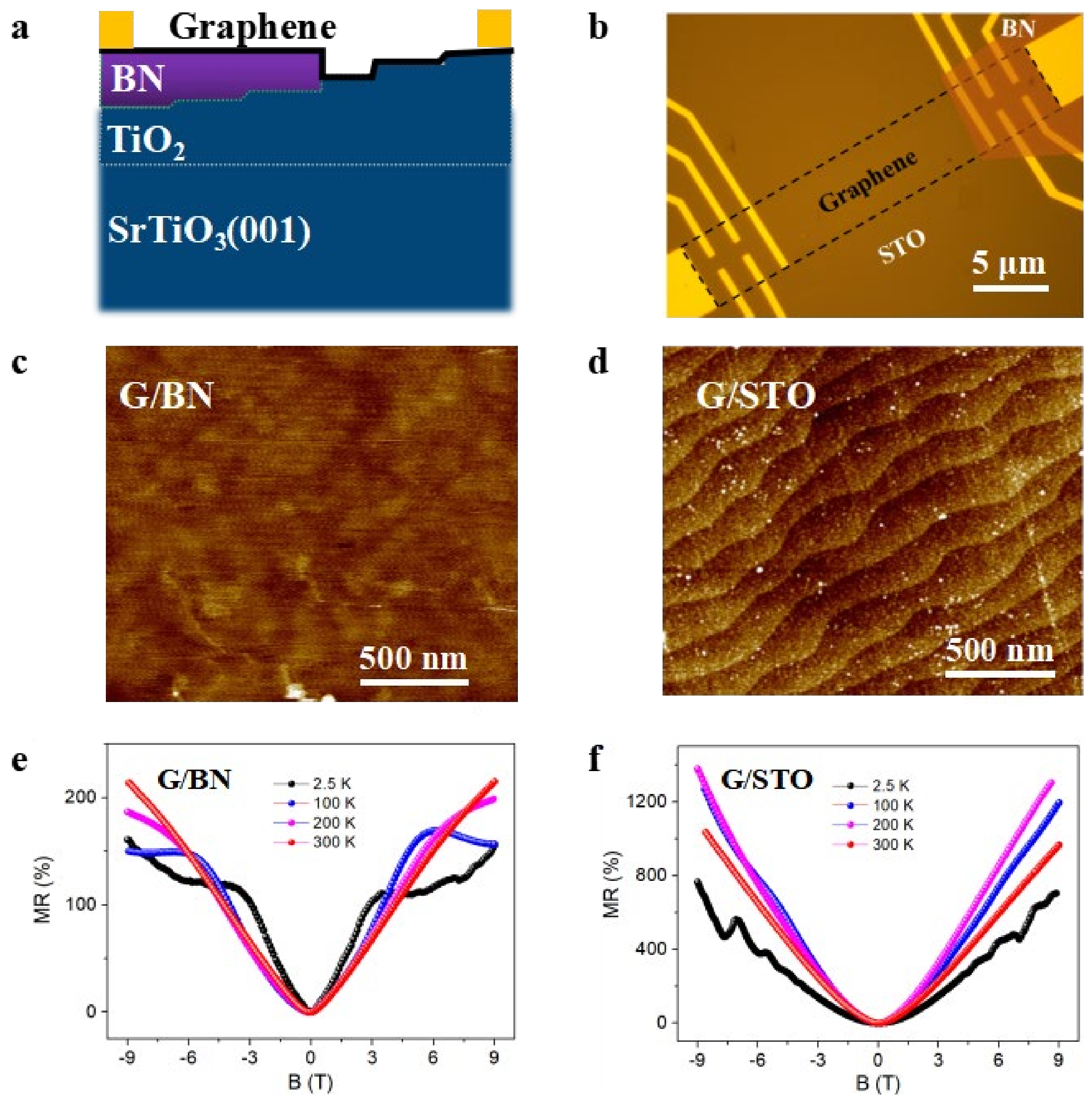

**Supplementary Figure S7 | MR of single-layer graphene on flat BN and terraced STO. a,** Schematic illustration of graphene on flat BN and terraced STO. The left side is graphene on flat BN and the right side is the graphene on terraced STO. The thickness of BN is 20-30 nm, thus the atomic steps of terraced STO can be totally screened by the thick BN. **b,** Optical micrograph of the fabricated Hall bar structure of graphene on flat BN and terraced STO. The area of graphene is marked by the dark dashed line. **c,d,** AFM images of graphene on flat BN and terraced STO, respectively. On the BN side, the graphene is flat while graphene on the terraced STO side adapts a terraced structure. **e,f,** Temperature-dependent MR of graphene on flat BN and terraced STO, respectively. At 2.5 K, both graphene on flat BN and terraced STO show the SdH oscillations with a positive MR background. At 300 K, the MR of graphene on flat BN is only 220%, in agreement with previous reported single-layer graphene on BN. On the contrary, the MR of graphene on terraced STO can be large as 1,020%, indicating the crucial role of terraces and steps of terraced STO in enhancing MR.

## S8. MR of terraced single-layer graphene on $TiO_2$-terminated BTO

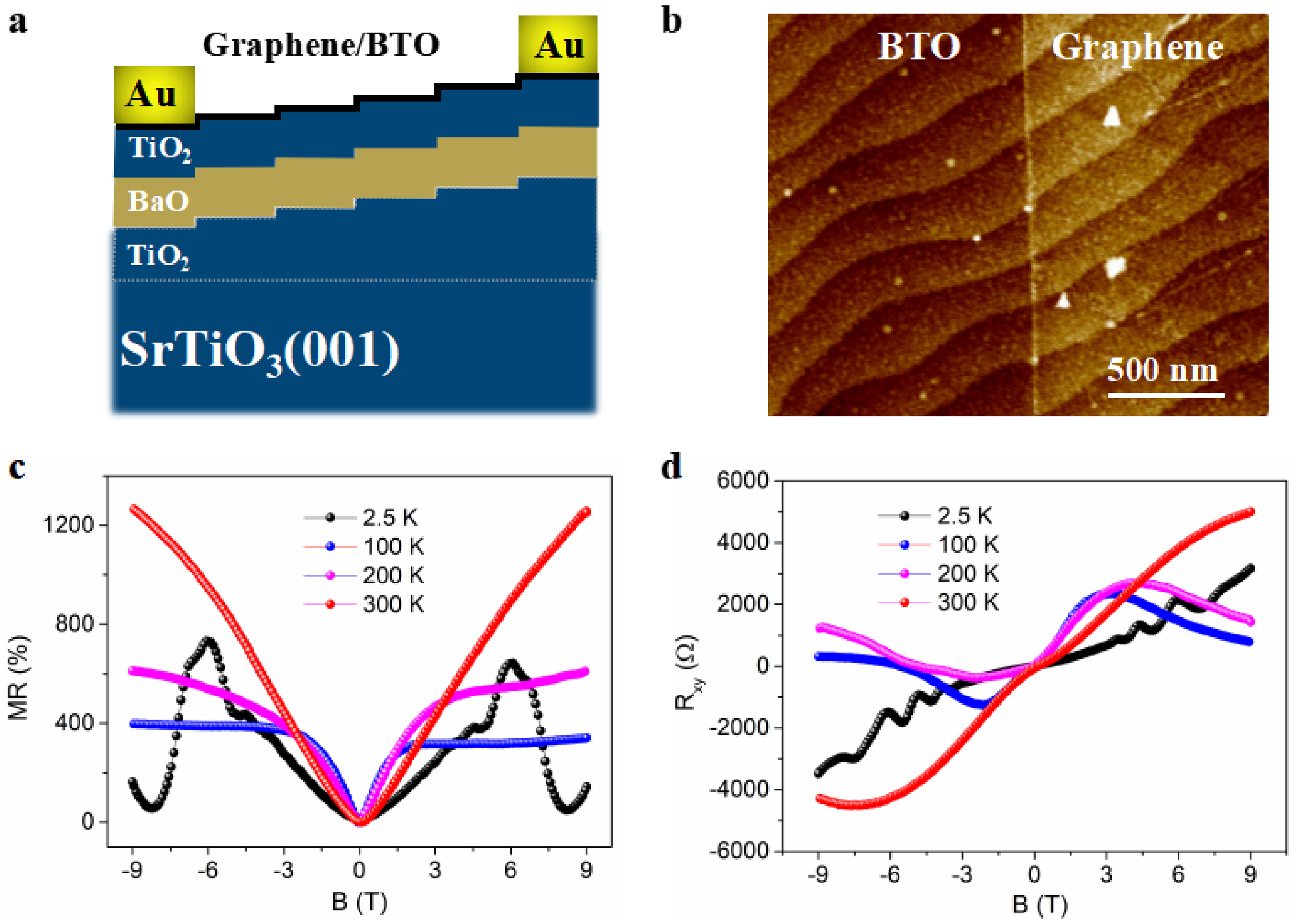

**Supplementary Figure S8 | MR of terraced single-layer graphene on $TiO_2$-terminated BTO. a,** Schematic illustration of terraced graphene on BTO. Due to the layer-by-layer growth mode, the graphene will form an interface with $TiO_2$ termination. **b,** AFM image of graphene on BTO. The left side is BTO and the right side is graphene on BTO. After the lamination of graphene on the terraced BTO, the surface of graphene shows similar terrace and sharp step features as those of BTO. **c,d,** Temperature-dependent MR and Hall of graphene on BTO. At low temperature (2.5 K) and under a strong magnetic field, the quantized Landau levels make the MR departing from linear dependence. The appearance of SdH oscillations make the MR at high field rather small, only 50%, which is different from the MR of graphene on other terminations like G/STO. The possible reasons can be the random topographic corrugation and charge puddles distribution induced by various terminations. At low field (0 - 1 T), the MR of G/BTO is roughly temperature independent. At high fields, the MR tends to saturate at 100-200 K, and finally linearly increases without saturation at 9 T. At room temperature, the MR of terraced graphene/BTO is as large as 1200% at 9 T, and the corresponding carrier density is $1.2 \times 10^{12}$ $cm^{-2}$.

**S9. MR of terraced single-layer graphene on $AlO_2$-terminated LAO**

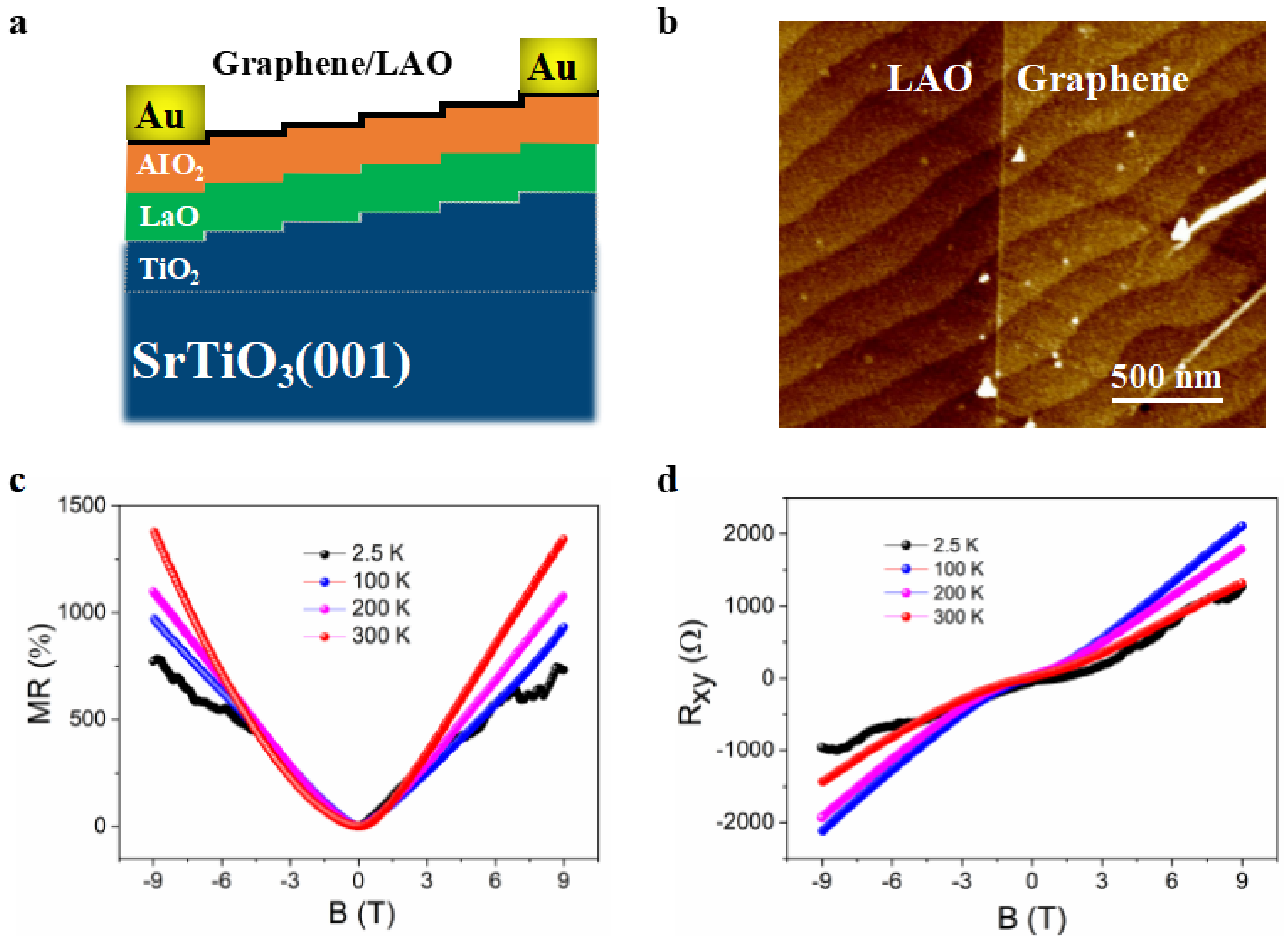

**Supplementary Figure S9 | MR of terraced single-layer graphene on $AlO_2$-terminated LAO. a,** Schematic illustration of terraced graphene on LAO. Due to the layer-by-layer growth mode, the graphene will form an interface with $AlO_2$ termination. Since there is 2D liquid gas at the interface of LAO/STO and to ensure that graphene is the only conducting channel, after the growth of LAO on the STO substrate, the LAO/STO is annealed at a high oxygen environment to remove the oxygen vacancy in STO. **b,** AFM image of graphene on LAO. The left side is LAO and the right side is graphene on LAO. After the lamination of graphene on the terraced LAO, the surface of graphene shows similar terrace and sharp step features as those of LAO. **c,d,** Temperature-dependent MR and Hall of graphene on LAO. At low temperature (2.5 K) and under a strong magnetic field, the quantized Landau levels make the MR of G/LAO departing from linear dependence and the appearance of SdH oscillations. However, the MR can still keep large, nearly 750%, at high field, which is similar to MR of G/STO. At field (0 - 4 T), the MR of G/BTO is roughly temperature independent. At high fields, the MR tends to be different at 100-200 K, and finally linearly increases without the signature of saturation at 9 T. At room temperature, the MR of terraced graphene/LAO is as large as 1,300% at 9 T, and the corresponding carrier density is $8.1 \times 10^{12}$ $cm^{-2}$.

**S10. MR of terraced single-layer graphene on $FeO_2$-terminated LFO**

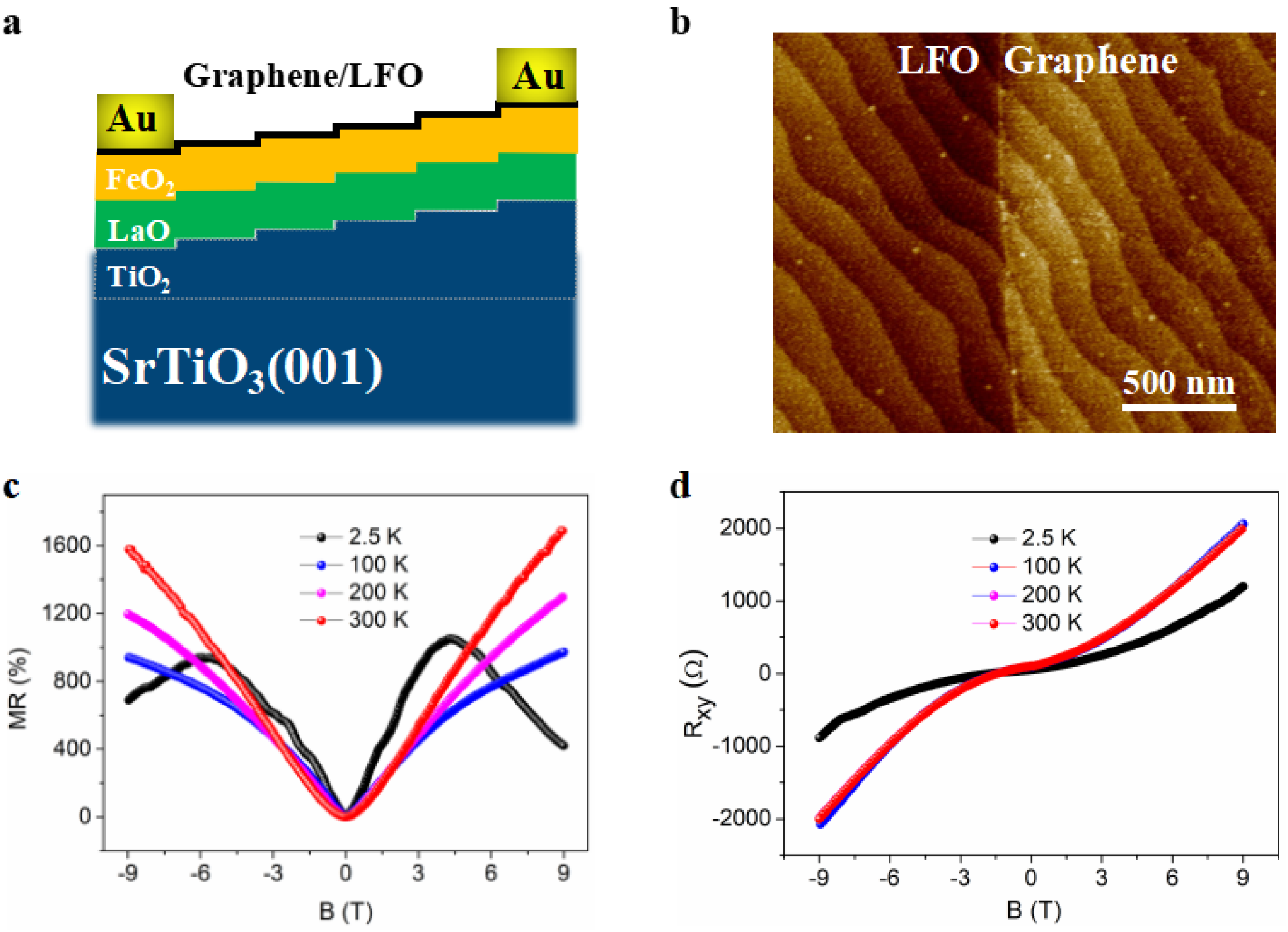

**Supplementary Figure S10 | MR of terraced single-layer graphene on $FeO_2$-terminated LFO. a,** Schematic illustration of terraced graphene on LFO. Due to the layer-by-layer growth mode, the graphene will form an interface with $FeO_2$ termination. **b,** AFM image of graphene on LFO. The left side is LFO and the right side is graphene on LFO. After the lamination of graphene on the terraced LFO, the surface of graphene shows similar terrace and sharp step features as those of LFO. **c,d,** Temperature-dependent MR and Hall of graphene on LFO. At low temperature (2.5 K) and under a strong magnetic field, the quantized Landau levels make the MR of G/LFO departing from linear dependence. The appearance of SdH oscillations makes the MR decrease at the high field. At field (0 - 3 T), the MR of G/LFO is roughly temperature independent. At high fields, the MR tends to be different and show the signature of saturation at 100-200 K, and finally linearly increases without saturation at 9 T. At room temperature, the MR of terraced graphene/LFO is as large as 1,660% at 9 T, and the corresponding carrier density is $5\times10^{12}$ $cm^{-2}$.

**S11. MR of terraced single-layer graphene on AlN-terminated AlN**

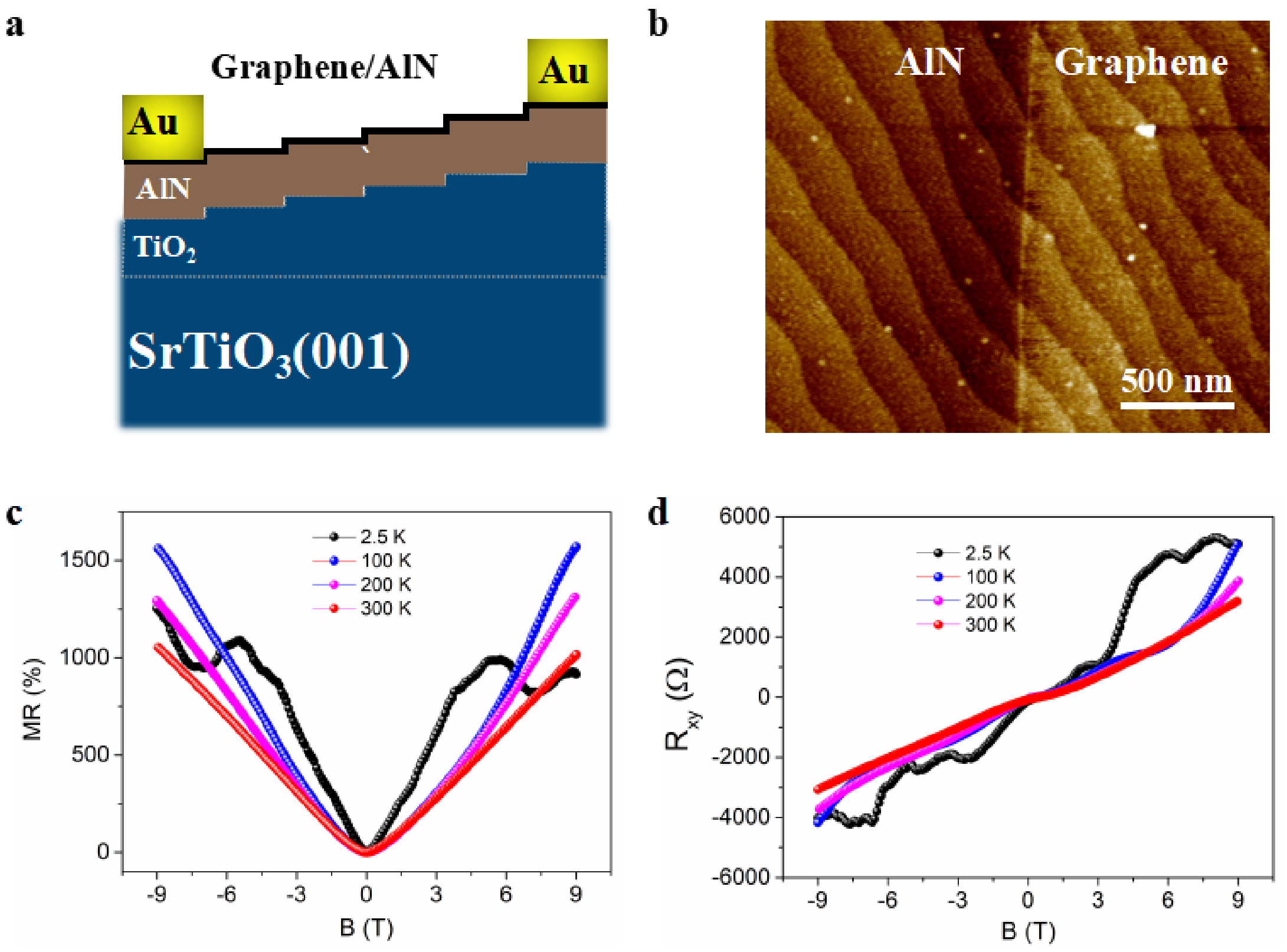

**Supplementary Figure S11 | MR of terraced single-layer graphene on AlN -terminated AlN. a,** Schematic illustration of terraced graphene on AlN. Different from the single-crystal BTO, LAO and LFO thin films grown at high temperature, the amorphous AlN thin film was grown at room temperature and also follows the terraced structure of the STO substrate. **b,** AFM image of graphene on AlN. The left side is AlN and the right side is graphene on AlN. After the lamination of graphene on the terraced AlN, the surface of graphene shows similar terrace and sharp step features as those of AlN. **c,d,** Temperature-dependent MR and Hall of graphene on AlN. At low temperature (2.5 K) and under a strong magnetic field, the quantized Landau levels make the MR of G/AlN departing from linear dependence. The appearance of SdH oscillations makes the MR saturate at nearly 1000% at high field, which is different from the MR of graphene on other terminations. At field (0-4 T), the MR of G/AlN is roughly temperature independent. At high fields, the MR tends to be different at 100-200 K, and finally linearly increases without the signature of saturation at 9 T. At room temperature, the MR of terraced graphene/AlN is as large as 1,060% at 9 T, and the corresponding carrier density is $1.8\times10^{12}$ $cm^{-2}$.

## S12 | Comparison of MR of graphene on flat BN and terraced BN

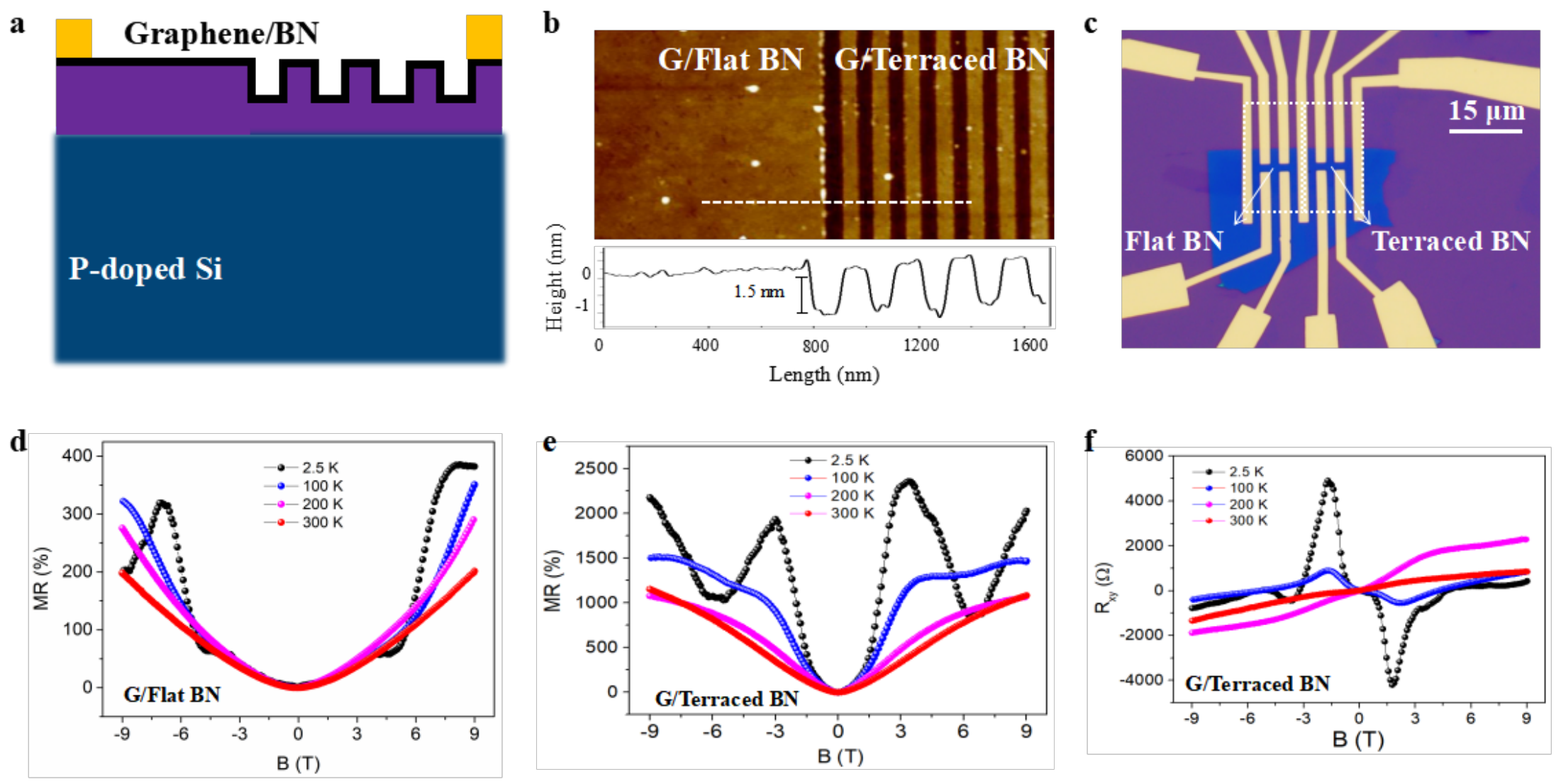

**Supplementary Figure S12 | MR of single-layer graphene on flat BN and terraced BN. a,** Schematic illustration of graphene on flat BN and terraced BN. The left side is graphene on flat BN and the right side is graphene on terraced BN. The thickness of BN is 10-20 nm. **b,** AFM images of graphene on flat BN (left side) and terraced BN (right side), respectively. On the flat BN, the profile of graphene is flat while graphene on terraced BN becomes the terraced structure with a height step of 1.5 nm and a width of 100 nm. **c,** Optical micrograph of the fabricated Hall bar structure of graphene on flat BN and terraced BN. The left side is the Hall bar of graphene on flat BN and the right side is the Hall bar of graphene on the terraced BN. The location of graphene is marked by the black dashed line. **d,e,** Temperature-dependent MR of graphene on flat BN and terraced BN, respectively. At 2.5 K, both graphene on flat BN and terraced BN show the SdH oscillations with a positive MR background. At 300 K and 9 T, the MR of graphene on flat BN is only 200%. On the contrary, the MR of graphene on the terraced BN can be as large as 1100%, indicating the crucial role of the terraces and steps of BN in enhancing the magnitude of the MR. **f,** Temperature-dependent Hall of graphene on the terraced BN. The carrier density is $3.65\times10^{12}$ cm$^{-2}$ at 300 K, estimated from the slope of the low field of Hall resistance.

The relatively high concentration in graphene on terraced BN can be attributed to the surface dangling bonds, which will introduce more charge carriers in graphene. Moreover, the trap air between graphene and terraced substrates can also be regarded a doping source. Similar with the case of graphene on terraced STO, there is also a sign change in Hall signal of graphene on terraced BN (Fig. S12f). From the band structure point of view, the origin of the temperature-dependent majority carriers is the shift of the Fermi-level, as discussed in Supplementary Figure S13, but more intrinsically, the temperature dependence behavior depends strongly on the amount of disorder. This is because at finite temperature, both electron and hole carriers will be activated, which will strongly influenced by the disorder-induced regions of inhomogeneous carrier density. This temperature dependence implies that the charge impurity scattering dominated the electrical transport of the terraced graphene from low to high temperature.

**S13. Temperature-dependent bipolar carrier tuning at the CNP and universal conductance fluctuation behavior at low-temperature transport**

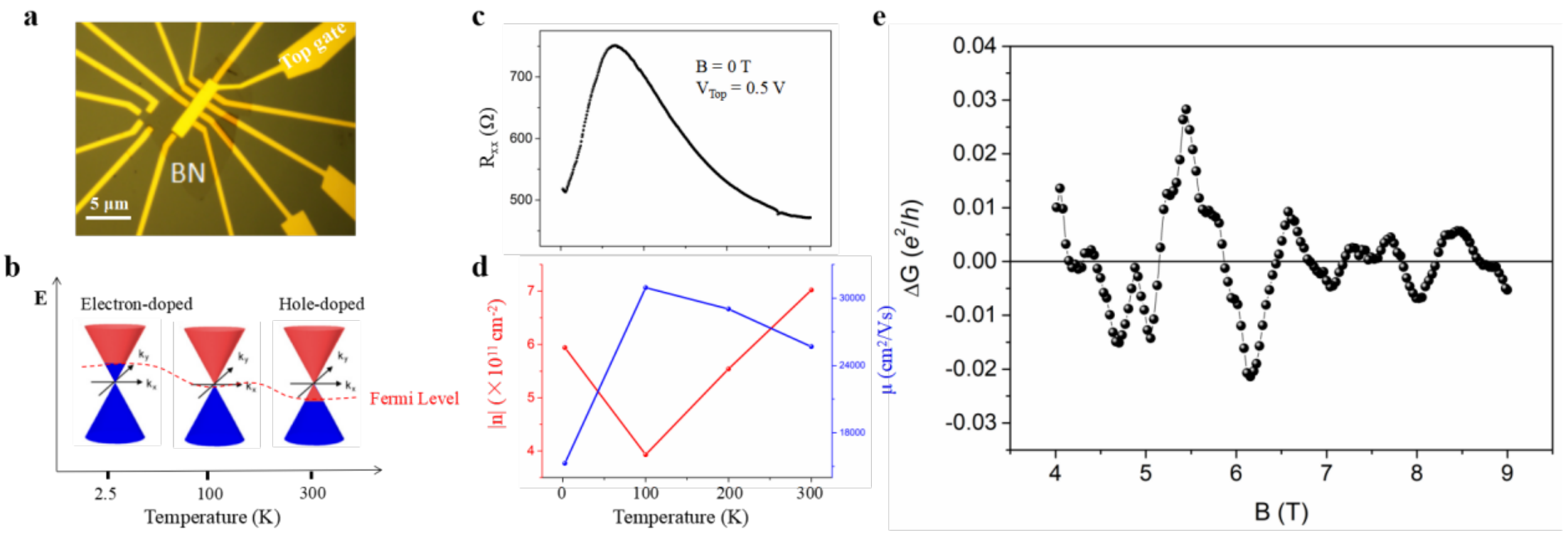

**Supplementary Figure S13 | Temperature-dependent transport properties of terraced single-layer graphene at the CNP and universal conductance fluctuation behavior at low-temperature transport. a,** Optical image of top-gate terraced graphene device. **b,** A sketch illustrating the shift of the Fermi level from high-temperature to low-temperature. **c,** Zero-field resistance $R_{xx}$ (B=0) with ramping the temperature from 300 K to 2 K. The resistance increase from 345 Ω as the sample is cooled from 300 K, peaks at 750 Ω near 80 K, and then quickly decreases to 520 Ω at 2.5 K. This behavior has been reproduced for both warm-up and cool-down cycles, with a minor shift. **d,** Extracted carrier density n and Hall mobility μ at different temperatures. Carrier density was estimated from the slope of the Hall coefficient. The observed resistance (Fig. S13c) is consistent with temperature-dependent transport parameters such as carrier density and Hall mobility (Fig. S13d). The carrier density falls to the lowest value of $n = 3.2\times10^{11}$ $cm^{-2}$ at round 100 K where the zero-field resistance peaks to its maximum. Meanwhile, the Hall mobility increases from $2.6\times10^4$ $cm^2/Vs$ at 300 K to a maximum value of $\mu = 3.2\times10^4$ $cm^2/Vs$ at 100 K and again decreases to a value of $\mu = 1.6\times10^4$ $cm^2/Vs$ at 2.5 K, indicating that the change of temperature can also tune the graphene Fermi level across the CNP, similar to the effect of electrical-field tuning of the Fermi-level. **e,** Universal conductance fluctuation at 2.5 K. The parabolic background was subtracted from the measured magnetoconductance from 4 to 9 T (the MR curve is shown in Fig. 3e) and obtained magnetoconductance fluctuations as shown in the Figure S13e. The magnetoconductance fluctuation is shown in $e^2/h$ units. It can be seen that the amplitude of conductance fluctuation is around 0.03 $e^2/h$. Moreover, the fluctuation shows the aperiodicity as a function of the magnetic field. These criteria suggest the universal conductance fluctuation (UCF) in our terraced graphene sample.

## S14. Strain analysis of terraced single-layer graphene at the nanoscale

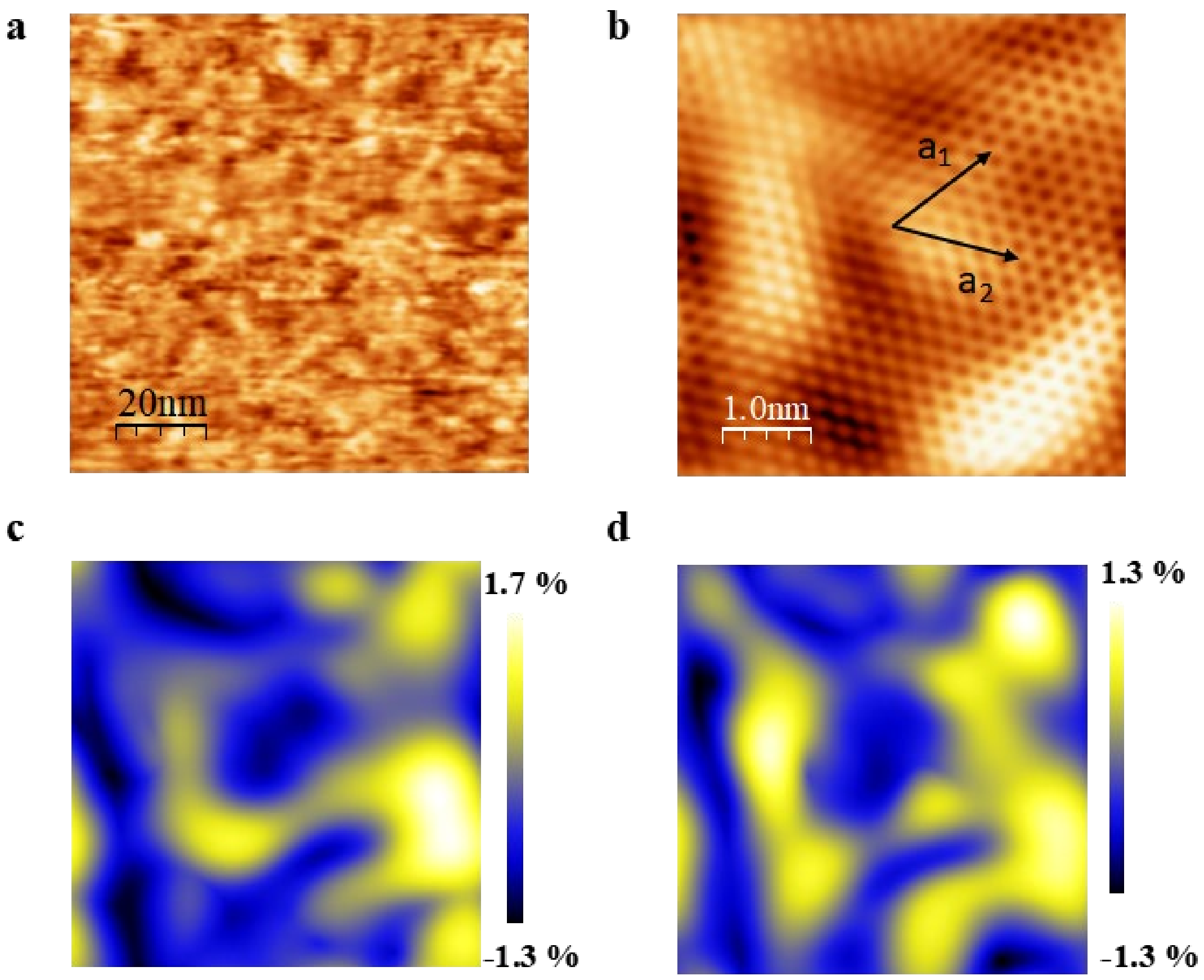

**Supplementary Figure S14 | Large area image, atomic resolution image and strain mapping of terraced graphene on STO. a,** Large area STM image of graphene on STO terrace (setpoint: $V_s$ = 500 mV, $I_s$ = 100 pA). **b**, Atomic resolution image of the wrinkles in graphene (setpoint: $V_s$ = -300 mV, $I_s$ = 50 pA). **c,d,** Strain maps along $a_1$ (**c**) and $a_2$ (**d**) directions as marked in (**b**). Yellow and blue represent the positive and negative lattice displacements. The scale bars have been converted into the strain percentage by d/a, where d is the measured displacement and a is the lattice constant of graphene, 246 pm.

Figure S14a shows graphene on the STO terrace containing dense bump and depression areas on the surface. Atomic resolution AFM (Fig. 4b in the main text) and STM (Fig. S14b) images also show subtle wrinkles, with which surface lattice distortion can be observed unambiguously. Meanwhile, to determine the in-plane strain with the picometer accuracy, we use the Lawler-Fujita algorithm to extract the lattice displacement[54]. Due to the small size of the STM image (5 nm) we choose a small coarsening length of 5.6 Å ($L = 1/\Lambda$), corresponding displacement mapping along two lattice directions $a_1$ and $a_2$ are extracted as shown in Fig. S14c and d, respectively. From the calculations, the maximum strain also can be derived as 1.7 %. Note that the 0.2% strain obtained from the geometrical model and Raman study in Fig. S2 is an average value providing estimation for the strain in the whole laminated graphene area. But in reality, there would be strain fluctuation in the terraced graphene.[34] The 1.7% strain observed in STM (Fig. S14) and 6.9% strain calculated in DFT are on the other hand the extracted maximum values that can demonstrate a correlation between the strain and charge density fluctuation in graphene.

## S15. Charge puddles of terraced single-layer graphene

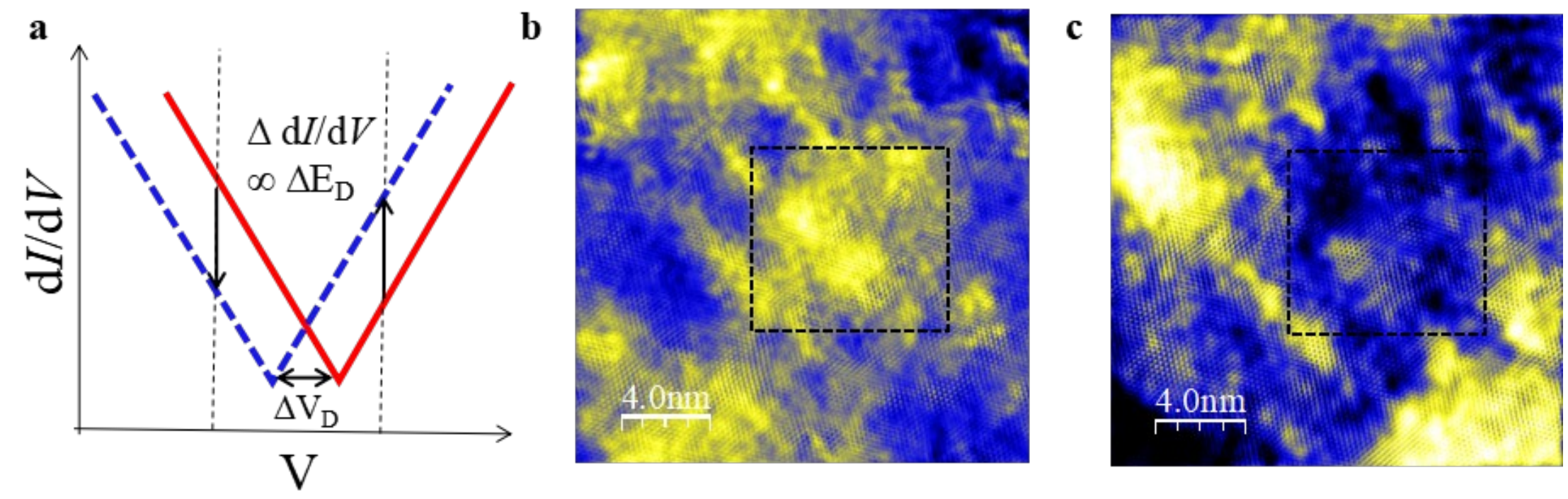

**Supplementary Figure S15 | Charge puddles in graphene. a,** Sketch shows how the changes in the Dirac-point ($\Delta V_D$) are proportionally related to the changes in d$I$/d$V$ (density of states (DOS)) intensity (Δ d$I$/d$V$) at fixed energy (vertical dashed lines). The d$I$/d$V$ signal intensity will be inversed at the energy above and below Dirac point. **b,c,** d$I$/d$V$ mapping got at the energy below Dirac point (**b**, $V_S$ = -0.2 V, $I_S$ = 200 pA) and above Dirac point (**c**, $V_S$ = 0.1 V, $I_S$ = 200 pA), respectively. Dashed square highlights a typical DOS inversed area.

The corrugations in the STM image (Fig. 4a in the main text) contain the information of both atomic structure and electronic structure. As shown in Fig.4b in the main text, nc-AFM can be used to detect the surface atomic corrugation directly. The correlated distribution of bump and depression areas with the STM image indicates the strain related atomic corrugation is the main contribution to the surface disorder. On the other hand, the differences between STM and nc-AFM images also imply the existence of surface electronic fluctuation. From the STS measurement at different sites, the Dirac point movement can be distinguished unambiguously (Fig. 4c in the main text), which can be ascribed to the in surface impurity doping induced electron puddle. As localized electron doping can be regarded as locally back gating, thus surface electron puddles can be visualized by mapping the spatial dependent Dirac point shift. Meanwhile, direct dI/dV mapping is another alternative approach. Considering the linear density of states (DOS) of graphene, electron doping induced Dirac point shift (left direction in Fig. S15a) will cause the DOS reduction at a fixed energy below the Dirac point and the DOS increase at a fixed energy above the Dirac point (e.g. the two energy as marked by two dashed lines in Fig. S15a). Therefore, d$I$/d$V$ (DOS) mapping on the electron doping will detect a lower conductance on the left side of Dirac point and inversed higher conductance on the right. Thus by measuring below and above the Dirac point, the electron puddles can be visualized in the two intensity inversed d$I$/d$V$ mapping (Fig. S15b,c), indicating the in-surface electronic disorder.

## S16. DFT calculations of graphene on STO and BN

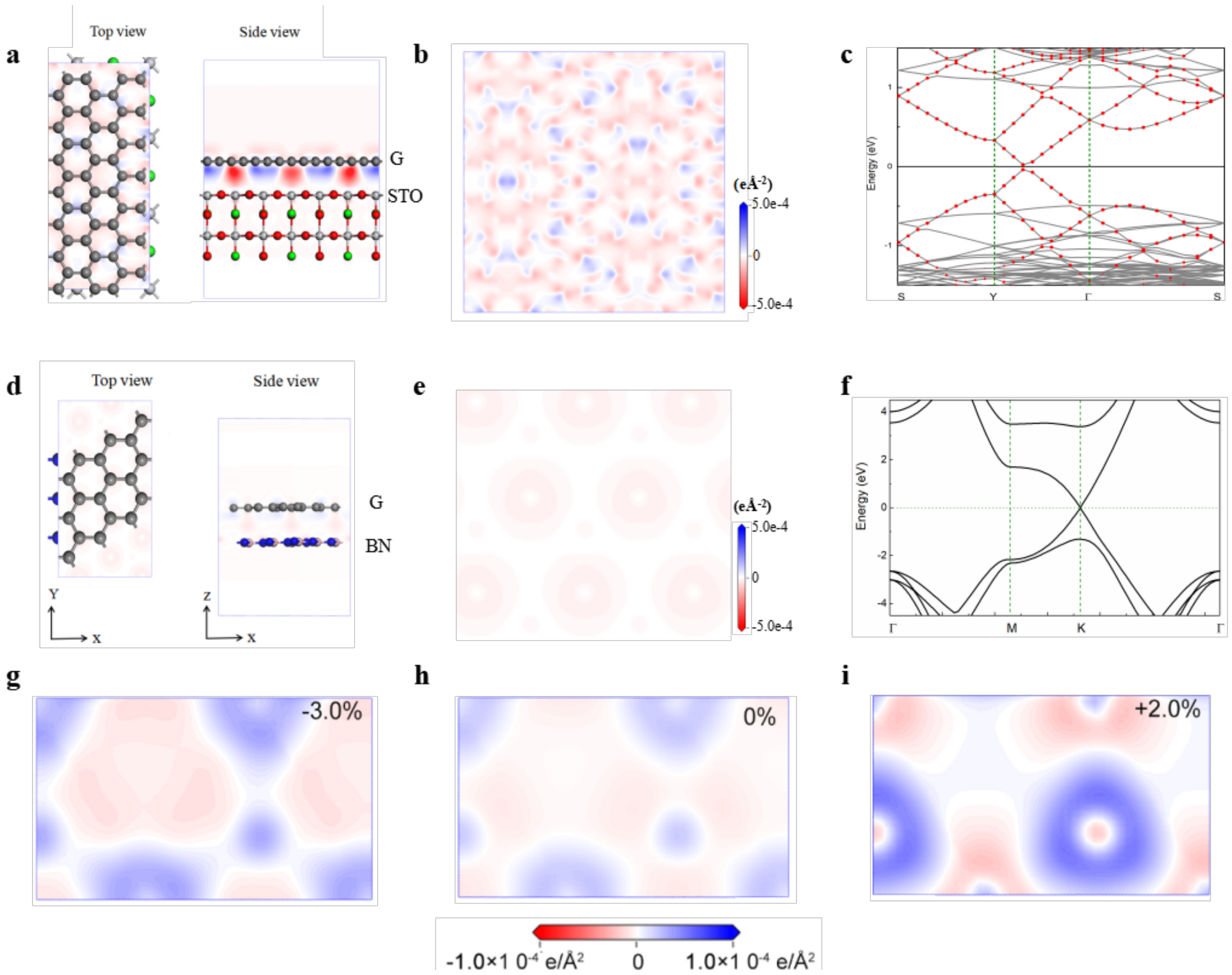

**Supplementary Figure S16 | DFT calculation of graphene on STO and BN. a,** Top view (left) and side view (right) of the optimized graphene on STO with a smaller strain (S-G/STO). Compared with that of L-G/STO, the basal plane of graphene remains flat on the STO. **b,** Top view of differential charge density of S-G/STO (projected on the graphene plane), which is visualized by an iso-surface value of $5 \times 10^{-4}$ eÅ$^{-2}$. **c,** Projected band structure of the graphene on STO, in which the dotted solid circle denotes the contribution of graphene. **d,** Top view (left) and side view (right) of optimized graphene on BN. **e,** Top view differential charge density of graphene/BN (projected on graphene plane), showing a more homogeneous charge density distribution. **f,** The band structure of the graphene on BN. The contour plot of Charge redistribution of graphene/BN heterostructure under strain: **g,** 3% compressive strain on graphene, **h,** no strain on graphene, and **i,** 2% tensile strain on graphene, in which the contour plot of differential charge density is projected on the graphene plane with an iso-surface value of $1.0 \times 10^{-4}$ e/Å$^2$, and the red and blue colour denotes the charge depletion and accumulation, respectively.

We also study the interface interaction of graphene/STO with smaller strain on the graphene (S-G/STO). As Fig. S16a shows, the basal plane of graphene remains flat on the STO, accompanied with a lager interfacial spacing of 3.2 Å due to the smaller lattice strain, in contrast to those of graphene/STO with larger strain on the graphene as discussed above. We also note a much reduced inhomogeneous charge redistribution in this S-G/STO as shown in Fig. S16b. The band structure of this interface is presented in Fig. S16c, from which the Dirac cone of graphene (solid red circles in the figure) can still be seen. Combined with results of the graphene/STO with large strain on graphene (L-G/STO), we can see that different regions of graphene may be affected by different compressive strains. For larger strain on graphene (L-G/STO), the basal plane of graphene becomes non-uniform, forming a warped structure, which leads to pronounced inhomogeneous electron-hole puddles in the graphene. While for a smaller strained graphene (S-G/STO), the basal plane of graphene remains flat, accompanied by a much less inhomogeneous charge redistribution. In both cases, the electronic structure of graphene is not changed significantly as evidenced by the well preserved Dirac cone. This indicates that the electron mobility of graphene remains high on the STO substrate. In contrast, the interaction between graphene and h-BN substrate is much weaker, as suggested by the nearly intact structure of graphene (see Fig. S16a), unnoticeable charge redistribution in the graphene (see Fig. S16b), and the perfect Dirac cone in the band structure (see. Fig. S16c). However, if we consider the charge density distribution of graphene/h-BN under strain and compare with the graphene without any external strain (Figure S16h), both compressive and tensile strain on graphene can introduce more pronounced inhomogeneous charge density distribution in graphene, as shown in Figure S16 g and i.

These results might be useful to understand the large MR observed in the experiment. The stronger interaction and shorter spacing between graphene and STO might facilitate the coupling between STO and graphene more effectively, which leads to stronger scattering due to electron-hole puddles in graphene and the phonon scattering from the substrate. On the other hand, the interfacial interaction between graphene and STO is not that strong enough to destroy the Dirac cone of graphene, so the high electron mobility of graphene does not decrease much. In addition, the high dielectric constant of STO also helps to screen the charged impurity scattering. Therefore, the stronger interfacial coupling, high-mobility in graphene, and large dielectric constant of STO lead to the observed the large MR of graphene/STO.

**S17. Comparison of absorption energy of graphene on STO, BN, and BP**

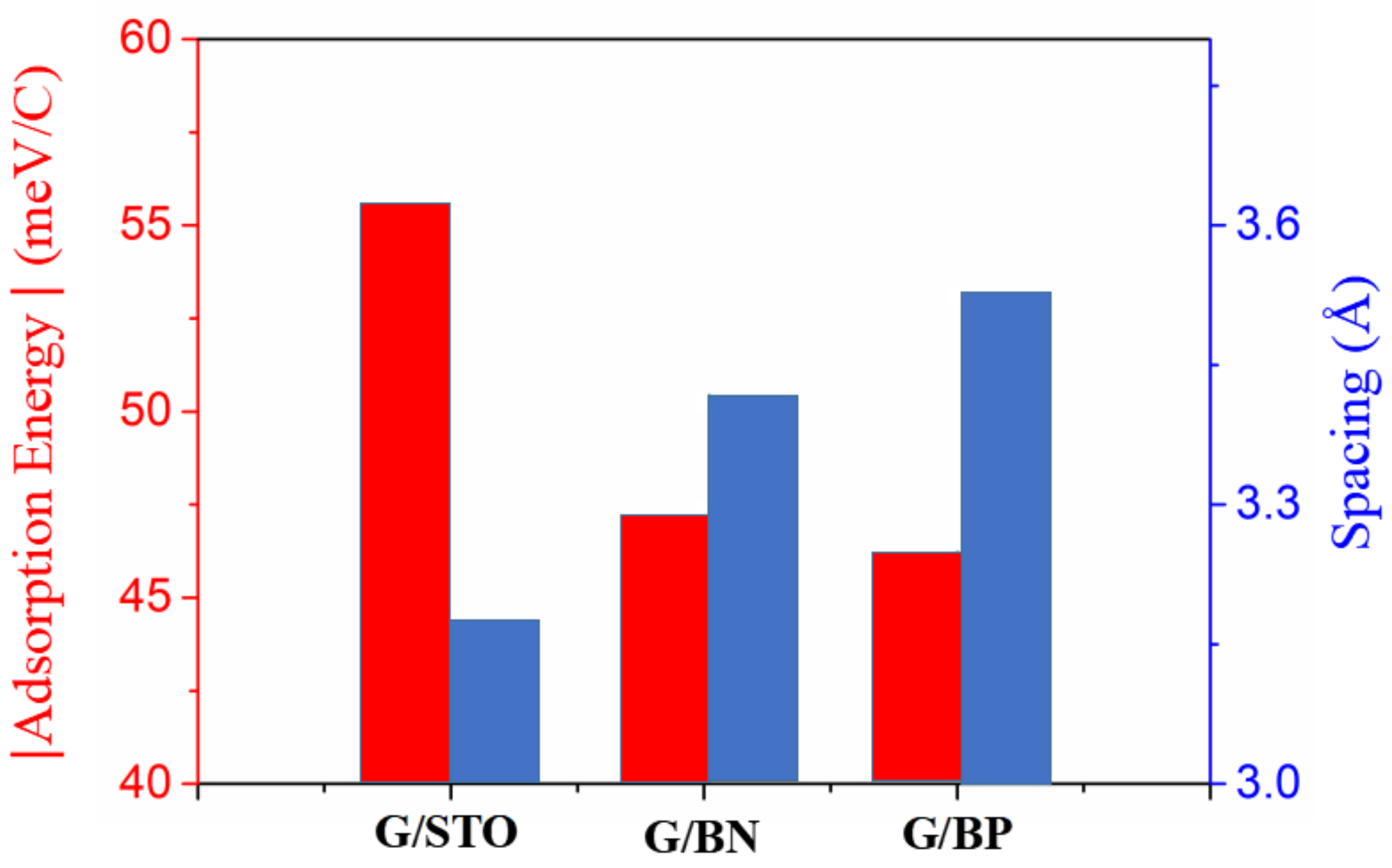

**Supplementary Figure S17 | Comparison of absorption energy and interfacial spacing for graphene on STO, BN, and BP substrates.**

The interfacial interaction strength can be estimated from the corresponding adsorption energy, in which the higher adsorption energy indicates stronger interface interaction. Figure S17 shows the comparison of interfacial interaction of graphene on various substrates such as h-BN, BP and STO. As we can see, the adsorption energy of graphene on STO is the highest, compared with BN and BP. From Fig. S17, we can also find that interfacial spacing between graphene and STO is also much smaller than that on BN and BP. Even for the S-G/STO, the interfacial spacing is about 3.2 Å (for L-G/STO, it is about 2.6 Å). The stronger interfacial interaction and smaller interfacial spacing might facilitate the coupling between graphene and STO, leading to stronger interfacial scattering.

## S18. Theoretical understanding of the disorder-induced MR in terraced single-layer graphene

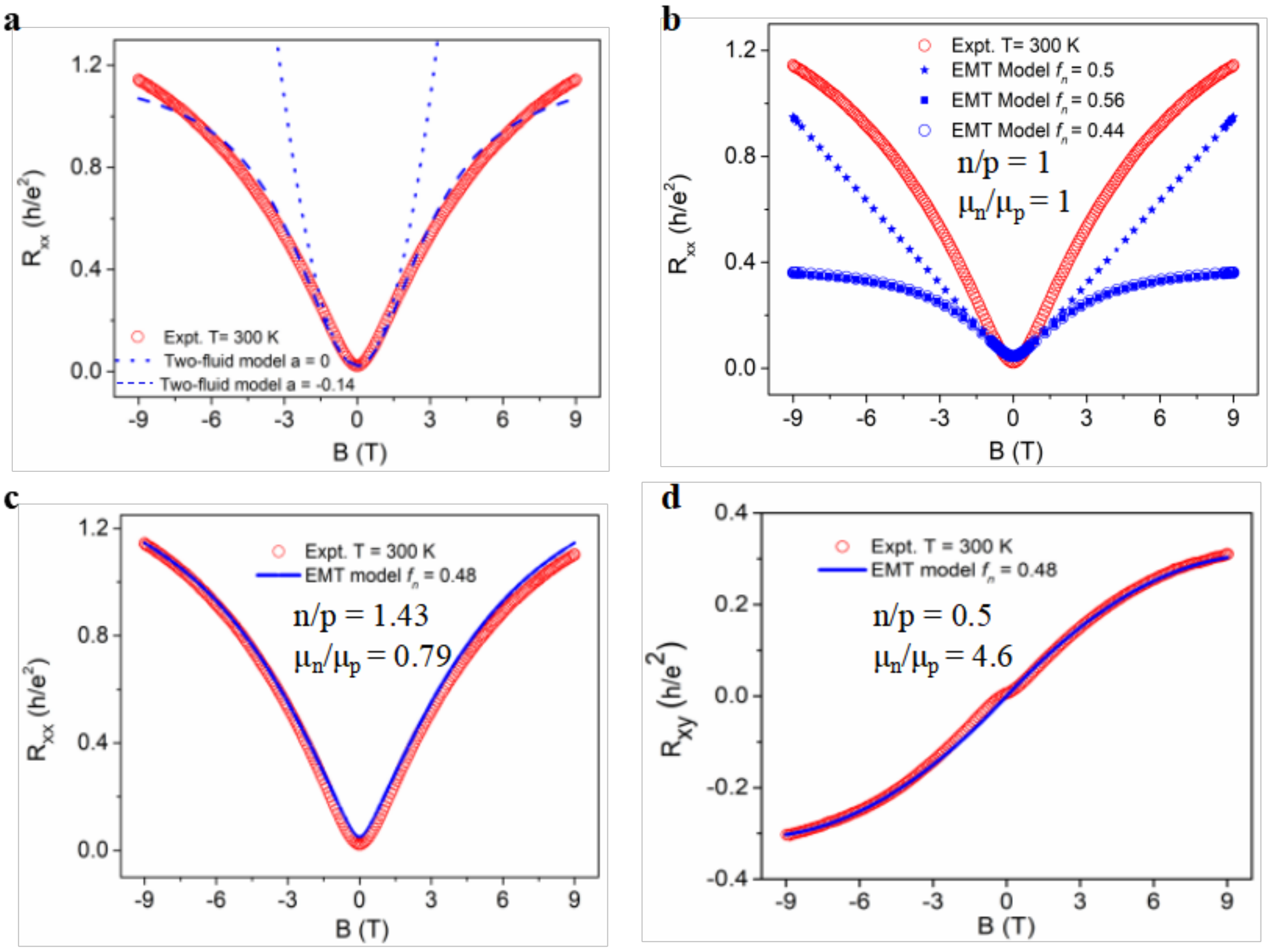

**Supplementary Figure S18 | MR and Hall fitting based on the two-fluid model and the effective medium model. a,** $R_{\mathrm{xx}}$ as a function of $B$ based on the two-fluid model fits. Open red line: Experimental data. Blue dot line is a fit to the two-fluid model with a = 0, and the blue dashed line is a fit to the two-fluid model with a=-0.14 and μ = 2.3 × $10^4$ $cm^2$/Vs. **b,** $R_{\mathrm{xx}}$ as a function of $B$ based on the EMT fit, assuming electrons and holes with equal mobility and equal concentrations, $n = p = 2.3 \times 10^{11}$ $cm^{-2}$ and $\mu_n = \mu_p = 2.3 \times 10^4$ $cm^2$/Vs. Star blue line: Calculated results with $f_n = 0.5$. Solid square blue line: calculated results with $f_n = 0.56$. Open circle blue line: $f_n = 0.44$. **c,** $R_{\mathrm{xx}}$ as a function of $B$ based on the effective medium model, assuming electrons and holes with different mobility and different concentrations. Solid blue line: $n = 1.6 \times 10^{11}$ $cm^{-2}$ and $p = 1.12 \times 10^{11}$ $cm^{-2}$. $\mu_n = 2.15 \times 10^4$ $cm^2$/Vs and $\mu_p = 2.71 \times 10^4$ $cm^2$/Vs. $f_n = 0.48$. **d,** $R_{xy}$ as a function of $B$ based on the effective medium model, assuming electrons and holes with different mobility and different concentrations. Solid blue line: $n = 2.3 \times 10^{11}$ $cm^{-2}$ and $p = 4.5 \times 10^{11}$ $cm^{-2}$. $\mu_n = 3.68 \times 10^4$ $cm^2$/Vs and $\mu_p = 0.8 \times 10^4$ $cm^2$/Vs. $f_n = 0.485$.

Firstly, we start from a homogeneous medium model, called the two-fluid model. A 2D conductor with a single carrier type exhibits no transverse $R_{\mathrm{xy}}(B)$ because the force exerted by the Hall field cancels the Lorentz force, the longitudinal MR is proportional to $1+(\mu B)^2$. However, for graphene, at the CNP, the existence of both electron and hole carriers can give rise to finite MR. Hwang *et al.*[55] proposed a two-fluid model where the resistivity $\rho_{xx}=\frac{w}{l}R_{\mathrm{xx}}$ is given by:

$$\rho_{xx}\ (B)=\ \rho_{xx}\ (0)\frac{1+(\mu B)^2}{1+(\alpha\mu B)^2} \tag{S5}$$

where *w* and *l* is the width length of the channel (in the experiment) and $\alpha=(n-p)\ /\ (n+p)$ with the concentrations of electrons and holes. This model can qualitatively explain the sharp peak in $\rho_{\mathrm{xx}}(B)$ at the CNP. At the CNP, $n=p$, thus $\alpha=0$, then $\rho(B)\propto(\mu B)^2$. While far away from the CNP, $\alpha\ =\ 1$, then $\rho(B)\ =\ \rho(0)$, which means the MR is zero. Figure S18a fits the $R_{\mathrm{xx}}$ based on the two-fluid model. When $\alpha=0$, a large linear MR is predicted. By choosing $\mu$ and $\alpha$ to fit our MR data, this theory can successfully fit the MR at a low magnetic field of 3 T. However, the two-fluid model predicts a saturation at high field, which is not in agreement with our MR data, suggesting that the two-fluid model is inadequate to accurately explain the MR.

Then, we go to the effective medium theory (EMT)[44], which is a model that describes graphene as a mixture of *n*- and *p*-type puddles. This model is reasonable if the *n*- and *p*-type puddles are distributed randomly and each puddle has its own conductivity. For tensor conductivity, the defining equation for the EMT is:

$$\sum_{i=(n,p)} f_i\,\delta\sigma_i(I-\Gamma\delta\sigma_i)^{-1}=0 \tag{S6}$$

where $f_n$ and $f_p$ are the area fraction of *n*- and *p*-type puddles and $f_n+\ f_p\ =1$. The

$$\Gamma=-\frac{I}{2\sigma_{e,xx}} \tag{S7}$$

is depolarization tensor for the planar geometry and *I* is the 2×2 unit matrix, and the

$$\delta\sigma_i=\sigma_i-\sigma_e \tag{S8}$$

with $\sigma_i$ is $\sigma_n$ and $\sigma_p$. $\sigma_e$ is the effective conductivity tensor of a graphene sheet which breaks into *n*- and *p*-type puddles and has four components:$\sigma_{e,xx}$, $\sigma_{e,xy}$, $\sigma_{e,yx}$ and $\sigma_{e,yy}$, where:

$$\sigma_n = \sigma_{n,0} \begin{pmatrix} \frac{1}{1+(\omega_{c,n}\tau_n)^2} & \frac{\omega_{c,n}\tau_n}{1+(\omega_{c,n}\tau_n)^2} \\ \frac{-\omega_{c,n}\tau_n}{1+(\omega_{c,n}\tau_n)^2} & \frac{1}{1+(\omega_{c,n}\tau_n)^2} \end{pmatrix} \quad (S9a)$$

$$\sigma_p = \sigma_{p,0} \begin{pmatrix} \frac{1}{1+(\omega_{c,p}\tau_p)^2} & \frac{\omega_{c,p}\tau_p}{1+(\omega_{c,p}\tau_p)^2} \\ \frac{-\omega_{c,p}\tau_p}{1+(\omega_{c,p}\tau_p)^2} & \frac{1}{1+(\omega_{c,p}\tau_p)^2} \end{pmatrix} \quad (S9b)$$

with $\sigma_{n,0}$ and $\sigma_{p,0}$ the zero-field conductivities $\omega_{c,n}$ and $\omega_{c,p}$ the cyclotron frequencies, and $\tau_n$ and $\tau_p$ the relaxation time of *n*- and $p$-type puddles.

We will calculate the effective conductivity tensor of the graphene sheet which are divided into *n*- and $p$-type puddles. After putting those terms in the equation S6, we have two coupled scalar equations for the two independent components of $\delta_{e,xx}$ and $\delta_{e,xy}$.

$$k(\sigma_{n,0}\sigma_{p,0}A_nA_p + \sigma_{n,0}A_n\sigma_{e,xx} - \sigma_{p,0}A_p\sigma_{e,xx} - \sigma^2{}_{e,xx} + \sigma_{n,0}\sigma_{p,0}B_nB_p + \sigma_{n,0}B_n\sigma_{e,xy} - \sigma_{p,0}B_p\sigma_{e,xy} - \sigma^2{}_{e,xy})$$
$$= (\sigma_{n,0}\sigma_{p,0}A_nA_p + \sigma_{p,0}A_p\sigma_{e,xx} - \sigma_{n,0}A_n\sigma_{e,xx} - \sigma^2{}_{e,xx} + \sigma_{n,0}\sigma_{p,0}B_nB_p - \sigma_{p,0}B_p\sigma_{e,xy} + \sigma_{n,0}B_n\sigma_{e,xy} - \sigma^2{}_{e,xy}) \quad (S10a)$$

$$k(\sigma_{n,0}\sigma_{p,0}B_nB_p + \sigma_{n,0}B_n\sigma_{e,xy} - \sigma_{p,0}B_p\sigma_{e,xy} - \sigma^2{}_{e,xy} + \sigma_{n,0}\sigma_{p,0}A_nA_p + \sigma_{n,0}A_n\sigma_{e,xx} - \sigma_{p,0}A_p\sigma_{e,xx} - \sigma^2{}_{e,xx})$$
$$= (\sigma_{n,0}\sigma_{p,0}B_pB_n - \sigma_{p,0}B_p\sigma_{e,xy} + \sigma_{n,0}B_n\sigma_{e,xy} - \sigma^2{}_{e,xy} + \sigma_{p,0}\sigma_{n,0}A_pA_n + \sigma_{p,0}A_p\sigma_{e,xx} - \sigma_{n,0}A_n\sigma_{e,xx} - \sigma^2{}_{e,xx}) \quad (S10b)$$

$$k(-\sigma_{n,0}\sigma_{p,0}A_nB_p - \sigma_{n,0}A_n\sigma_{e,xy} + \sigma_{p,0}B_p\sigma_{e,xx} + \sigma_{n,0}\sigma_{p,0}B_nA_p + \sigma_{n,0}B_n\sigma_{e,xx} - \sigma_{p,0}A_p\sigma_{e,xy})$$
$$= (\sigma_{n,0}\sigma_{p,0}A_pB_n - \sigma_{p,0}B_p\sigma_{e,xy} - \sigma_{n,0}B_n\sigma_{e,xx} - \sigma_{p,0}\sigma_{n,0}B_pA_n - \sigma_{p,0}B_p\sigma_{e,xx} -$$
$$\sigma_{n,0}A_n\sigma_{e,xy} \quad (S10c)$$

$$k(-\sigma_{n,0}\sigma_{p,0}B_nA_p - \sigma_{n,0}B_n\sigma_{e,xx} + \sigma_{p,0}A_p\sigma_{e,xx} + \sigma_{n,0}\sigma_{p,0}A_nB_p + \sigma_{n,0}A_n\sigma_{e,xy} - \sigma_{p,0}B_p\sigma_{e,xx})$$
$$= (\sigma_{n,0}\sigma_{p,0}A_nB_p + \sigma_{p,0}B_p\sigma_{e,xx} + \sigma_{n,0}A_n\sigma_{e,xy} - \sigma_{n,0}\sigma_{p,0}B_nA_p + \sigma_{p,0}A_p\sigma_{e,xy}$$
$$+ \sigma_{n,0}B_n\sigma_{e,xx} \quad (S10d)$$

where $= -f_n/f_p$ ; $\sigma_{n,0} = ne\mu_n$ ; $\sigma_{p,0} = pe\mu_p$ ; $A_n = 1/(1+(\mu_nB)^2)$ ; $A_p = 1/(1+(\mu_pB)^2)$; $B_n = \mu_nBA_n$; $B_p = \mu_pBA_p$

We are using these equation $(S10\ a, b, c, d)$ to fit our experimental data $R_{xx}$ and $R_{xy}$ as shown in Figure S18. From the fitting, we have calculated 7 derived parameters, $n,\ p,\ \mu_n,\ \mu_p,\ B,\ \mathfrak{f}_n,\ \mathfrak{f}_p$.

The effective resistivity tensor is then obtained by inverting the matrix $\delta_e$ so that

$$\rho_{\mathrm{e,xx}} = \frac{\sigma_{e,xx}}{\left({\sigma^2}_{e,xx} + {\sigma^2}_{e,xy}\right)} \qquad (S11a)$$

$$\rho_{\mathrm{e,xy}} = -\frac{\sigma_{e,xy}}{\left({\sigma^2}_{e,xx} + {\sigma^2}_{e,xy}\right)} \qquad (S11b)$$

We can calculate the longitudinal and Hall resistance from the solutions of $\delta_{e,xx}$ and $\delta_{e,xy}$. First, we consider a special case, $n = p$ and $\mu_n = \mu_p$, then we can get:

$$(k-1)\left(A_0^2\sigma_0^2 + B_0^2\sigma_0^2 - {\sigma^2}_{e,xx} - {\sigma^2}_{e,xy}\right) = 0 \qquad \text{(S12a)}$$

$$(1+k)B_0\sigma_{e,xx} + (1-k)A_0\sigma_{e,xy} = 0 \qquad \text{(S12b)}$$

where $A_0 = \ 1/(1+(\mu B)^2);\ B_0 = \ \mu B/(1+(\mu B)^2)$.

Solving the above two equations, we can get:

$$\sigma_{e,xx} = \sqrt{\frac{A_0^4\sigma_0^2(k-1)^2 + A_0^2B_0^2\sigma_0^2(k-1)^2}{A_0^2(k-1)^2 + B_0^2(k+1)^2}} \qquad \text{(S13a)}$$

$$\sigma_{e,xy} = \frac{(1+k)B_0}{(k-1)A_0}\sigma_{e,xx} \qquad \text{(S13b)}$$

If we consider a special situation, $k = -1$, then $\mathfrak{f}_n = \mathfrak{f}_p = 0.5$, which means that graphene has equal $n$ and $p$-type puddles, then we can get:

$$\rho_{xx}(B) = \left[\frac{\delta_{xx,0}}{[1+(\mu B)^2]^{1/2}}\right]^{-1} \qquad \text{(S14)}$$

Where $\sigma_{xx,0}$ is the conductivity at zero fields. This is the MR of inhomogeneous electrons and holes with equal mobility and equal carrier concentrations. However, no matter, how to tune the parameters, the fit is not good. At $k = -1$, the $R_{xx}$ varies exactly linear with the magnetic field. At other puddle fractions, it quickly saturates, as shown in Fig. S18b.

Finally, we consider the situation that carrier density and mobility are not equal for electrons and holes. We assure : $n = 1.6 \times 10^{11}$ cm$^{-2}$ ; $p = 1.12 \times 10^{11}$ cm$^{-2}$ and $\mu_n = 2.15 \times 10^4$ cm$^2$/Vs; $\mu_p = 2.71 \times 10^4$ cm$^2$/Vs. When the $n$-type puddle fraction is 0.48, as can be seen in Fig. S18c, the fit to the experimental data is excellent over most of the field range, especially at high fields, indicating the EMT model is superior to the two-fluid model. Moreover, the Hall data can also be well fit, by ensuring that the fraction of electron puddles is 0.485 and $n$ = 2.3$\times 10^{11}$ cm$^{-2}$; $p$ = 4.5$\times 10^{11}$ cm$^{-2}$; $\mu_n$ = 3.68 × $10^4$ cm$^2$/Vs; $\mu_p$ = 0.8 × $10^4$ cm$^2$/Vs.

Here we use different parameters to fit the MR and Hall data. When the same fitting parameters were used for both $R_{xx}$ and $R_{xy}$ data, we found that the $R_{xy}$ data can only be fit well by the EMT at a low field regime of up to 2 T. The discrepancy at higher fields is due to the assumptions made in the model. First, the model assumes that the puddles have a circular form and the medium surrounding the puddles is uniform. However, in our samples, the puddles are of a random and complicated distribution. Second, the topographic corrugation induced by the step of the terraced substrates in our samples represents a network of disorders with a random nature which will strongly affect the scattering of the charge carriers under a magnetic field. The EMT model used is not able to consider such a particular network of disorders. The influence of these two assumptions (the geometry of the puddles and the exact network of disorders) is not dominant and thus leads to a better fit at low fields. However, at high fields, they start to play an important role, resulting in the deviation of the fits if we use the same parameters for both $R_{xx}$ and $R_{xy}$.

Moreover, the EMT model is based on two-fluid carriers and only accurates close to CNP. At higher carrier concentration, the large MR in terraced graphene can be qualitatively understood by the disorder-induced multiple conduction channels in an inhomogeneous medium system. Because of the disorder induced by the terraced substrate, the charge carriers can be considered to be travelling in a non-uniform medium consisting of different conducting channels. The fluctuations of carrier density and mobility in these different conducting channels can lead to a large MR effect even at a high carrier density.[56] To simplify this multiple-channel model, we consider two channels as an example. If the $\mu B > 1$, the MR can be simplified as the following[24]:

$$MR = \frac{n_1 n_2}{(n_1 + n_2)^2} \frac{(\mu_1 - \mu_2)^2}{\mu_1 \mu_2} \qquad (S15)$$

where $\mu_1$ and $n_1$ are the mobility and carrier density of the channel 1 and $\mu_2$ and $n_2$ are the mobility and carrier density of the channel 2. It can be seen that if the system is very homogeneous (for example when $n_1 = n_2$ and $\mu_1 = \mu_2$), the MR will then be approaching zero. Our terraced graphene is a very inhomogeneous system consisting of many different channels with different mobilities. Assuming a two-channel system with a large difference in mobility or density, we can see from the equation that the MR can remain arbitrarily large."

**Table 1 | Summary of room-temperature local MR of graphene between this work and literature**

| Graphene Layers | Substrates | T (K) | B (T) | MR (%) | References |
|---|---|---|---|---|---|
| 1 | $SiO_2$ | 300 | 9 | 140 - 275 | *Carbon* **136,** 211 (2018)<br>*Nat.Commun.* **4,** 1921 (2013)<br>*Nat.Commun.* **6,** 8337 (2015)<br>*Nano Lett.* **18,** 3377 (2018)<br>*PRB* **88,** 195429 (2013) |
| 1 | BN | 300 | 9 | 180 - 200 | *Science* **357,** 181 (2017) |
| 1 | SiC | 300 | 9 | 60-80 | *Materials Letters* **174,** 118 (2016)<br>*Nano Lett.* **10,** 3962 (2010)<br>*Appl. Phys. Lett.* **105,** 182102 (2014) |
| 1 | BP | 300 | 9 | 775 | *Nano Lett.* **18,** 3377 (2018) |
| 1 | Au-coating<br>N-doping<br>Fluorinating | 300 | 9 | -4 - 480 | *APL* **105,** 143103 (2014)<br>*ACS Nano* **9,** 1360 (2015)<br>*PRB* **83,** 085410 (2011) |
| 1-2 | Au | 300 | 9 | 100 - 200 | *Nat.Commun.* **4,** 1921 (2013) |
| 4-6 | BN | 300 | 9 | 800 - 2000 | *Nat. Commun.* **6.**8337 (2015) |
| 1 | Terraced STO | 300 | 9 | 5000 | **This work** |

**Supplementary Table 1 | Room-temperature MR of graphene on various substrates.** The MR of terraced single-layer graphene is found to be as high as 5,000% at 9 T, which is one order of magnitude higher than previously reported single-layer

graphene devices at the same conditions.